\documentclass[11pt,english]{article}
\usepackage{amssymb}
\usepackage{mathrsfs, amsmath}
\usepackage[dvips]{graphicx}
\usepackage{setspace}
\usepackage{fancyhdr}
\usepackage{xcolor}
\usepackage{ifpdf}
\usepackage{graphicx}
\usepackage{rotating}
\usepackage{comment,tikz}
\usepackage{url}
\definecolor{darkblue}{rgb}{0,0,0.6}
\usepackage[colorlinks=true,citecolor=darkblue,linkcolor=black,urlcolor=darkblue]{hyperref}
\usepackage{cite}
\usetikzlibrary{patterns}

\makeatletter
\newcommand*{\defeq}{\mathrel{\rlap{%
                     \raisebox{0.3ex}{$\m@th\cdot$}}%
                     \raisebox{-0.3ex}{$\m@th\cdot$}}%
                     =}
\makeatother

\DeclareMathOperator{\p}{\partial}

\DeclareMathOperator{\Tr}{Tr}

\newcommand{\be}{\begin{equation}}
\newcommand{\ee}{\end{equation}}

\newcommand{\f}{\frac}
\newcommand{\tr}{\textrm{Tr}\,}

\newcommand{\va}{\textrm{vac}}
\newcommand{\ev}{\varepsilon_{\textrm{vac}}}

\begin{document}
\unitlength = 1mm
\ 

\begin{center}

{ \LARGE {\textsc{\begin{center}Emergent gravity from Eguchi-Kawai reduction \end{center}}}}

\vspace{0.8cm}
Edgar Shaghoulian

\vspace{.5cm}

{\it Department of Physics} \\
{\it University of California}\\
{\it Santa Barbara, CA 93106 USA}

\vspace{1.0cm}

\end{center}

\begin{abstract}

\noindent Holographic theories with a local gravitational dual have a number of striking features. Here I argue that many of these features are controlled by the Eguchi-Kawai mechanism, which is proposed to be a hallmark of such holographic theories. Higher-spin holographic duality is presented as a failure of the Eguchi-Kawai mechanism, and its restoration illustrates the deformation of higher-spin theory into a proper string theory with a local gravitational limit. AdS/CFT is used to provide a calculable extension of the Eguchi-Kawai mechanism to field theories on curved manifolds and thereby introduce ``topological volume independence." Finally, I discuss implications for a general understanding of the extensivity of the Bekenstein-Hawking-Wald entropy.
\end{abstract}

\pagebreak
\tableofcontents
\setcounter{page}{1}
\pagestyle{plain}

\setcounter{tocdepth}{1}
\section{Introduction}
Holographic theories with a local gravitational dual have several remarkable features that can be read off by analyzing (semi-)classical gravity in Anti-de Sitter space (AdS). To understand the emergence of gravity, it is important to understand precisely in the language of quantum field theory what mechanism is responsible for these features. Much of the work in this direction has focused on constraints from conformal field theory (CFT). Conformality is not an essential feature of holography. On the other hand, every holographic theory to date can be understood as a large-$N$ gauge theory. It is therefore natural to leverage whatever power such a structure brings us. This brings us to the idea of Eguchi-Kawai reduction.

The proposal of Eguchi and Kawai was that large-$N$ $SU(N)$ lattice gauge theory could be reduced to a matrix model living on a single site of the lattice \cite{Eguchi:1982nm}. This equivalence was postulated through an analysis of the Migdal-Makeenko loop equations (the Schwinger-Dyson equations for Wilson loop correlation functions) \cite{Makeenko:1979pb, Makeenko:1980vm} and assumed the preservation of center symmetry in the gauge theory. However, it was immediately noticed \cite{Bhanot:1982sh} that the center symmetry is spontaneously broken at weak coupling, disallowing the consistency of the reduction with a continuum limit. The authors of \cite{Bhanot:1982sh} further proposed the first in a long list of modifications to the gauge theory in an attempt to prevent center symmetry from spontaneously breaking. Their proposal is known as the quenched Eguchi-Kawai model, further studied in \cite{Gross:1982at}, where the eigenvalues of the link matrices were frozen to a center-symmetric distribution. Another proposed variant is known as the twisted Eguchi-Kawai model, wherein each plaquette in Wilson's action is ``twisted" (multiplied by) an element of the center of the gauge group \cite{GonzalezArroyo:1982hz}. Numerical studies have shown these early modifications fail at preserving center symmetry as well \cite{Bietenholz:2006cz, Teper:2006sp, Azeyanagi:2007su, Bringoltz:2008av}. 

Let us turn to the continuum. Whether or not center symmetry is preserved is often checked analytically by pushing the theory into a weakly coupled regime and calculating the one-loop Coleman-Weinberg potential for the Wilson loop around the compact direction. This is an order parameter for the center symmetry, and a nonvanishing value indicates a breaking of center symmetry. An early analytic calculation of the Coleman-Weinberg potential indicates the center-symmetry-breaking nature of Yang-Mills theories \cite{Gross:1980br}. Nevertheless, there are a few tricks that seem to work at suppressing any center-breaking phase transitions: a variant of the original twisted Eguchi-Kawai model \cite{GonzalezArroyo:2010ss}, deforming the action by particular double-trace terms \cite{Unsal:2008ch}, or considering adjoint fermions with periodic boundary conditions \cite{Kovtun:2007py}.  For a modern review see \cite{Lucini:2012gg}.

In this work, we will not be concerned with suppressing center-breaking phase transitions. Instead, we will focus on implications of the Eguchi-Kawai mechanism within center-symmetric phases. This will not be a restriction to the confined phase since we will be considering center symmetry with respect to spatial and thermal cycles. As we will be working in the continuum, let us formulate the continuum version of the Eguchi-Kawai mechanism. Consider a $d$-dimensional large-$N$ gauge theory compactified on $\mathcal{M}^{d-k}\times (\mathbb{S}^1)^k$ with center symmetry at the Lagrangian level. If translation symmetry and center symmetry are not spontaneously broken along a given $S^1$, then correlation functions of appropriate single-trace, gauge-invariant operators are independent of the size of that $S^1$ at leading order in $N$. We will review these notions in the rest of the introduction and spend section \ref{corr} elaborating on which sorts of observables are ``appropriate." This is often called large-$N$ volume independence, where ``volume" in particular refers to the size of the center-symmetric $S^1$s.

The Eguchi-Kawai mechanism is a robust, nonperturbative property of large-$N$ gauge theories that preserve certain symmetries. Famously, large-$N$ gauge theories also play a starring role in holographic duality. Curiously both contexts involve emergent spacetime in radically different ways. In this work we will be interested in what predictions the Eguchi-Kawai mechanism makes about gravity in AdS. Since the proposal concerns only leading-in-$N$ observables, we will be dealing exclusively with the (semi-)classical gravity limit in AdS.  
 A simple example illustrating the mechanism at work is the temperature-independence of the free energy density on $\mathcal{M}^{d-1}\times S^1_\beta$ at leading order in $N$ in the confined phase (e.g. $\mathcal{N}=4$ super Yang-Mills on $S^3_R\times S^1_\beta$). In AdS/CFT, this occurs because the thermal partition function is given by the contribution of thermal AdS below the Hawking-Page phase transition, whose on-shell action has an overall factor of inverse temperature $\beta$. When the theory deconfines, the free energy density becomes a nontrivial function of $\beta/R$. Physically, a hot gas of confined degrees of freedom like glueballs and mesons contributes a nontrivial function of $\beta/R$ to the free energy density, but it is subleading in $1/N$ compared to the vacuum contribution. To look ahead to another example of large-$N$ volume independence, see equation \eqref{toshow}.

We will spend the next section reviewing introductory material, ending with the central tool of this work, which is that a smooth, translation-invariant  gravitational description implies center symmetry preservation along all but one cycle. Center symmetry can spontaneously break along a given cycle as its size is varied, but there must only ever be one cycle which breaks the symmetry. We will refer to these transitions as center-symmetry-swapping transitions (CSSTs). The rest of the paper will leverage this structure to learn primarily about universal features of gravity, but also to learn about the Eguchi-Kawai mechanism in large-$N$ gauge theories. For some previous work exploring the Eguchi-Kawai mechanism in holography, see \cite{Furuuchi:2005qm, Furuuchi:2005eu, Poppitz:2010bt, Young:2014jma}.
\subsection{Summary of results}
Our primary tool will be that a smooth, translation-invariant gravitational description of a state or density matrix in a toroidally compactified CFT preserves center symmetry along all but one cycle. We will use this to produce the following universal features of gravity in AdS: (a) an extended range of validity of the general-dimensional Cardy formula, (b) the exact phase structure (including thermal and quantum phase transitions) with a toroidally compactified boundary, (c) a sparse spectrum of light states on the torus, (d) leading-in-$N$ connected correlators will be given by the method of images under smooth quotients of the spacetime, which reproduces the behavior of tree-level Witten diagrams, and (e) extensivity of the entropy for spherical/hyperbolic/planar black holes which dominate the canonical ensemble; for planar black holes this implies the Bekenstein-Hawking-Wald area law. (a)-(c) are closely related and can be found in section \ref{zeropoint}, (d) can be found in section \ref{corr}, and (e) can be found in section \ref{ext}. Using gravity to learn about the Eguchi-Kawai mechanism, we will find new center-stabilizing structures for strongly coupled holographic theories and propose an extension of the mechanism to curved backgrounds in section \ref{fieldtheory}.

\section{Center symmetry and Wilson loops}
Consider pure Yang-Mills theory on manifold $\mathcal{M}_{d-1} \times S_\beta^1$ with gauge group $G$ (for example $SU(N)$) with nontrivial center $C$ (for example $\mathbb{Z}_N$):
\be
S = -\f{1}{4} \int d^d x F_{\mu \nu}^a F^{a\mu\nu};\qquad F_{\mu \nu}^a = \p_\mu A_\nu^a - \p_\nu A_\mu^a+f^{abc}A_\mu^b A_\nu ^c\,.
\ee
This theory is invariant under the gauge symmetry
\be
A_\mu \rightarrow g A_\mu g^{-1}+g \p_\mu(g^{-1})
\ee
for $g: \mathcal{M}_{d-1} \times S_\beta^1\rightarrow G$ a map from our spacetime into the gauge group. The field strength transforms as 
\be
F_{\mu\nu}^a\rightarrow g F_{\mu \nu}^a g^{-1}\,.
\ee 
Let us consider the function $g$ to be periodic along the $S_\beta^1$ only up to an element of the gauge group: $g(x, \tau+\beta) =  g(x,\tau) h$ for $h \in G$. For $A_\mu$ to remain periodic we need $g(x,\tau+\beta) A_\mu(x,\tau+\beta) g^{-1}(x,\tau+\beta) = g(x,\tau)A_\mu(x,\tau) g^{-1}(x,\tau)$, in other words $A_\mu(x,\tau+\beta) = h^{-1} A_\mu(x,\tau) h$. But this requires $h \in C$ so we can commute it past $A_\mu$ and cancel it against $h^{-1}$. So we see that we can consistently maintain twisted gauge transformations as long as we twist by an element of the center. The action above is invariant under these extended gauge transformations. The space of physical states are constrained to be singlets under the usual gauge group $G$ but not under the twisted gauge transformations. In particular, Wilson loops which wrap an $S^1$, which will henceforth be referred to as Polyakov loops, transform under the generalized gauge transformation. To see this, consider the path-ordered exponential, i.e. the holonomy of the connection, around the $S^1$:
\be
\Omega_x(\tau+\beta, \tau)=P \exp\left[\int A_\mu dx^\mu\right]\longrightarrow g(x,\tau+\beta) \Omega_x(\tau+\beta,\tau)g(x,\tau)^{-1}\,.
\ee
The $P$ stands for path. We will refer to the trace of this object as the Polyakov loop, which for ordinary gauge transformations causes $g$ and $g^{-1}$ to annihilate by cylicity. For twisted gauge transformations, however, we are left with  
\be
W(C) \equiv \Tr\Omega_x(\tau+\beta,\tau)\longrightarrow h W(C)\,.
\ee
The $W$ stands for Polyakov. Thus the expectation value of a Polyakov loop can serve as an order parameter for the spontaneous breaking of center symmetry.

We will always take our trace in the fundamental representation, since the vanishing of the expectation value of such a loop is necessary and sufficient for the preservation of center symmetry, independent of the matter content. Contrast this with the case of rectangular Wilson loops (traces of path-ordered exponentials where the path traces out a large rectangle instead of wrapping an $S^1$) where the trace needs to be evaluated in the same representation as that of the matter content to access the energy required to deconfine the matter.

Let us now specify to gauge group $SU(N)$. The center of the gauge group is $\mathbb{Z}_N$, consisting of elements $z_n =\exp(2\pi i n/N)1\!\!1$. Every representation of $SU(N)$ can be classified by which of the $N$ representations of $\mathbb{Z}_N$ it falls under. This is called the $N$-ality of the representation, and it is determined by counting the number of boxes mod $N$ of the Young tableau of the representation. The addition of matter to our gauge theory explicitly breaks the center symmetry of the Lagrangian unless the matter is in a representation of vanishing $N$-ality \cite{greensite}. Fundamental representations have $N$-ality $1$ and therefore explicitly break center symmetry. Adjoint representations, on the other hand, have vanishing $N$-ality and therefore preserve center symmetry.

Even for matter in vectorlike representations that break center symmetry, there is an effective emergence of the symmetry as $N\rightarrow \infty$ as long as the number of vectorlike flavors is kept finite. This is simply because quarks decouple at leading order and one is left with the pure Yang-Mills theory. Interestingly, by orientifold dualities, even matrix representations (which break center symmetry and for which the matter does not decouple) have an emergent center symmetry at infinite $N$ \cite{Armoni:2007kd, Shifman:2007kt}.

\subsection*{Calculating Wilson loops in AdS}
There is a simple prescription for calculating the expectation value of a Wilson loop in the fundamental representation of the gauge theory using classical string theory. One calculates $e^{-S_{NG}}$ for the Nambu-Goto action $S_{NG}$ for a Euclidean string worldsheet which ends on the contour of the Wilson loop $\mathcal{C}$ \cite{Maldacena:1998im}. 
Let us specify to Polyakov loops wrapping an $S^1$ on the boundary. Notice that if this circle is non-contractible in the interior then we have $\langle W(\mathcal{C})\rangle = 0$. We now illustrate a famous example where this criterion distinguishes confined and deconfined phases. The thermally stable (i.e. large) AdS-Schwarzschild black hole, which has a thermal circle which caps off in the interior, admits a string worldsheet and therefore gives a nonvanishing Polyakov loop expectation value. This indicates a deconfined phase, which is appropriate as the AdS-Schwarzschild black hole is the correct background for the gauge theory at high temperature. Thermal global AdS, however, has a thermal circle which does not cap off in the interior and therefore gives a vanishing Polyakov loop expectation value.  This indicates a confined phase, which is appropriate for the theory at low temperature. Indeed, the bulk canonical phase structure for pure gravity indicates a transition between these two backgrounds when the inverse temperature is of order the size of the sphere. Similarly, the entropy transitions from $\mathcal{O}(1)$ in the confined phase (no black hole horizon) to $\mathcal{O}(N^2)$ in the deconfined phase (yes black hole horizon).

There is one more basic geometric fact we will need. Consider an asymptotically Euclidean AdS$_{d+1}$ spacetime with toroidal boundary conditions. Preserving translation invariance along the non-radial directions -- a necessary condition for the Eguchi-Kawai mechanism to work -- gives a metric of the form
\be
ds^2 = \f{dr^2}{r^2} + g_{\mu\nu}(r)d\phi^\mu d\phi^\nu\,,\qquad g_{\mu\nu}(r\rightarrow \infty) = r^2 \delta_{\mu\nu}\,.
\ee
To avoid conical singularities (e.g. metrics which look locally like $r^2(d\phi_1^2+ d\phi_2^2)$), no more than one of the boundary circles can cap off in the interior of the spacetime. While it may be possible that none of the boundary circles cap off in the interior (say through the internal manifold capping off instead), I do not know of any smooth, geodesically complete examples. We will therefore not consider this possibility, so in our context exactly one cycle caps off and the other $d-1$ circles remain finite-sized. This motivates the following simple yet extremely powerful statement. \emph{In any smooth, translation-invariant geometric description, the expectation value of Polyakov loops in the fundamental representation vanish in $d-1$ of the directions.} For theories with an explicit center symmetry, this means that we will have volume independence along $d-1$ directions as discussed in the introduction. Appropriate observables will therefore be independent of the sizes of the circles. For the gravitational description to be valid, the circles in the interior need to remain above string scale. For a translation of this criterion into field theory language, and in particular a discussion of Eguchi-Kawai reduction to zero size, see appendix \ref{circles}.

Just like the original Eguchi-Kawai example of pure Yang-Mills, our theory will of course deconfine, as signaled by the Hawking-Page phase transition in the bulk. This is sometimes called partial Eguchi-Kawai reduction, since the reduction only holds in the center-symmetric phase. We will refer to the ``Eguchi-Kawai mechanism" and ``large-$N$ volume independence" to describe this state of affairs.  (Large-$N$ volume independence refers in particular to independence of the size of center-symmetric $S^1$s, not necessarily the overall volume.) From our point of view, the deconfinement transition is just a center-symmetry-swapping transition (CSST) from the thermal cycle to a spatial cycle. It remains true that $d-1$ of the cycles preserve center symmetry. CSSTs can also occur between spatial cycles as they are varied. In this case, the transition is unrelated to confinement of degrees of freedom, since the entropy is $\mathcal{O}(1)$ before and after the transition. It instead signals a quantum phase transition, which can take place at zero temperature. Interestingly, this quantum phase transition persists up to a critical temperature.

\section{Reproducing gravitational phase structure/sparse spectra/extended range of validity of the Cardy formula}\label{zeropoint}
We will now show that the semiclassical phase structure of gravity in AdS is implied by our center symmetry structure. 
Consider an asymptotically AdS$_{d+1}$ spacetime with toroidal boundary conditions. The cycle lengths will be denoted $L_1,\dots, L_d$ with $\beta = L_1$.  We will pick thermal periodicity conditions for any bulk matter along all cycles and will comment at the end about different periodicity conditions. Assuming a smooth and translation-invariant description, the phase structure implied by gravity can succinctly be written in terms of the free energy density as 
\be\label{toshow}
f(L_1,\dots,L_d)\equiv-\f{\log Z(L_1,\dots, L_d)}{L_1 L_2\cdots L_d}=-\f{\ev}{L_{\text{min}}^{d}}\,,
\ee
where $\ev$ is a pure positive number (independent of any length scales) characterizing the vacuum energy on $S^1_L\times\mathbb{R}^{d-2}$ as $E_\va/V\equiv-\ev /L^d$ for spatial volume $V$ \cite{Shaghoulian:2015lcn}, and $L_{\text{min}}$ is the length of the smallest cycle. This is the phase structure \emph{independent} of the precise bulk theory of diffeomorphism-invariant gravity, as long as we maintain translation invariance and consider the thermal ensemble. Like in AdS$_3$, all the data about higher curvature terms is packaged into $\ev$.

Notice that the triviality of this phase structure implies highly unorthodox field theory behavior.  The phase structure \eqref{toshow} implies thermal phase transitions as the thermal cycle $\beta$ becomes the smallest cycle. There are also quantum phase transitions when two spatial cycles are smaller than the rest of the cycles (including $\beta$), and the larger of the two is changed to become smaller. These are quantum phase transitions because they can (and do) occur when $\beta \rightarrow \infty$, so they are not driven by thermal fluctuations. These quantum phase transitions, however, persist at finite temperature. Finally, in any given phase the functional form of the free energy density is independent of all cycle lengths except for one! Much of \cite{Belin:2016yll} was focused on reproducing this structure in field theory, and we refer the reader to that work to see the many nuances involved. 

We now turn to the gauge theory. We will see that our framework gives \eqref{toshow} immediately, thereby locating the points where phase transitions occur and the precise functional form of the free energy in all phases. Consider a field theory with our assumed center symmetry structure, which is that all but one cycle preserve center symmetry. We also have thermal periodicity conditions for the matter fields along all cycles, since this will give thermal periodicity conditions for the bulk matter fields and preserve modular $S$ invariance between any pair of cycles. Notice that by extensivity of the free energy and modular invariance \cite{Shaghoulian:2015lcn, Shaghoulian:2015kta}, we have 
\be\label{asymp}
f(L_1\rightarrow 0, L_2,\dots,L_d) = -\f{\ev}{L_1^{d}}\,.
\ee
Since the free energy density is supposed to be independent of the center-symmetry preserving directions, we deduce that the $L_1$ cycle breaks center symmetry. This is consistent with the expected deconfinement of the theory. Now let us consider varying any of the cycle sizes. As long as there is no center-symmetry-swapping transition (CSST), $f(L_1,\dots,L_d)$ continues to depend only on $L_1$. Since the theory is scale invariant, this fixes the $L_1$ dependence and we continue to have the behavior \eqref{asymp}. 
Finally, any CSST that occurs between cycle $L_\alpha$ and cycle $L_\gamma$ has to occur when $L_\alpha = L_\gamma$ by the modular symmetry between all cycles. 
So, when cycle lengths are equal, they must be symmetric: either they both preserve the center or they are undergoing a CSST. They cannot both break the center since only one cycle can ever break the center in our framework. 

Using the above facts that $f(L_1,\dots,L_d)$ can only change its functional form at CSSTs and that two cycles which have equal length must have the same center-symmetry structure, we can deduce the entire phase structure. As we increase $L_1$ from $L_1 = 0$, due to the symmetry between cycles there \emph{must} be a CSST between $L_1$ and the next-smallest cycle when they become equal. 
As $L_1$ is increased further, it is a center-preserving cycle passing other center-preserving cycles, so no more CSSTs can occur and the free energy density remains unchanged. Starting from an arbitrary torus, with an arbitrary cycle taken asymptotically small, this argument produces for us the entire phase structure \eqref{toshow}. 

What about the case where we do not preserve the symmetry between cycles? An interesting example of this is if we pick bulk fermions to be periodic along some cycles. In the gravitational picture these cycles are not allowed to cap off in the interior since this would not lead to a consistent spin structure. Thus, the phase structure is just as in \eqref{toshow}, where now $L_{\text{min}}$ minimizes only over the cycles with antiperiodic bulk fermions. We will comment more on the field theory implications of this in section \ref{percen}. To predict this bulk phase structure, we need to supplement our assumption of $d-1$ cycles preserving center symmetry with an assumption about which cycles preserve center symmetry for \emph{all} cycle sizes. These cycles can then never undergo CSSTs with other cycles. By repeating the arguments above, we can reproduce this modified bulk phase structure. 

\subsection{Extended range of validity of Cardy formula}
Holographic gauge theories, in addition to having the remarkable phase structure exhibited above, have an extended range of validity of the general-dimensional Cardy formula.  The Cardy formula in higher dimensions was derived in \cite{Shaghoulian:2015lcn, Shaghoulian:2015kta} and reproduces the entropy of toroidally compactified black branes at asymptotically high energy:
\be
\log \rho(E\rightarrow \infty) \approx \f{d}{(d-1)^{\f{d-1}{d}}}(\ev V_{d-1})^{\f{1}{d}}E^{\f{d-1}{d}}\,.
\ee
This precisely mimics how the two-dimensional Cardy formula \cite{Cardy:1986ie} reproduces the entropy of BTZ black holes at asymptotically high energy \cite{Strominger:1997eq}. Large $N$ operates as a thermodynamic limit that can transform our statements about the canonical partition function into the microcanonical density of states (this is discussed for example in the appendices of \cite{Hartman:2014oaa, Belin:2016yll}). We find that the Cardy formula is not valid only asymptotically, but instead is valid down to $E=-(d-1)E_\va$, which in canonical variables is at a symmetric point $\beta = L_{i,\textrm{min}}$ where $L_{i,\text{min}}$ is the smallest spatial cycle. This is precisely the energy at which the Hawking-Page phase transition between the toroidally compactified black brane and the toroidally compactified AdS soliton occurs in the bulk! Similar arguments in the case of non-conformal branes should give an extended range of validity for the Cardy formula of \cite{Shaghoulian:2015dwa}.

\subsection{Sparse spectra in holographic CFTs}
A sparse spectrum is often invoked as a fundamental requirement of holographic CFTs, and we have several avenues of thought that lead to this conclusion. Here we will be concerned with the sparseness necessary to reproduce the phase structure of gravity \cite{Hartman:2014oaa, Belin:2016yll}, not with the sparseness necessary to decouple higher-spin fields \cite{Heemskerk:2009pn}. 

We have already reproduced the complete phase structure \eqref{toshow}. By the arguments in \cite{Hartman:2014oaa, Belin:2016yll} this implies a sparse low-lying spectrum
\be
\rho(E<-(d-1)E_\va)\lesssim \exp\left(L_{i,\text{min}}(E-E_\va)\right),
\ee
where $L_{i,\text{min}}$ is the smallest spatial cycle. To roughly recap the argument of \cite{Belin:2016yll}, modular constraints on the vacuum energy coupled with the phase structure imply vacuum domination along all cycles except the smallest one. But to be vacuum dominated means that excited states do not contribute to the partition function. This leads to the constraint above, which is really a constraint on the entire spectrum, but is written as above since for $E>-(d-1)E_\va$ we have a precise functional form for the density of states: it takes the higher-dimensional Cardy form, which trivially satisfies the Hagedorn bound above. 

One can also access additional sparseness data by investigating different boundary conditions. To point out the simplest case, consider super Yang-Mills theory in a given number of dimensions with fermions having periodic boundary conditions along one cycle and antiperiodic boundary conditions along another cycle. Then modular covariance will equate a thermal partition function $Z_{NS,R}$ with a twisted partition function $Z_{R,NS}$ (twisted by $(-1)^F$), which will access the twisted density of states $\rho_B(E) - \rho_F(E)$. By similar steps as performed above, one will conclude a sparseness bound for this twisted density of states. The fact that preserving center symmetry can imply a supersymmetry-like bound is carefully discussed in a non-supersymmetric context in \cite{Basar:2013sza, Basar:2014jua}. 
\subsection{$SL(2,\mathbb{Z})$ family of black holes}

In this section and the next we will consider the case of twists between the cycles of the torus. We will begin with three bulk dimensions, where there is an extended family of solutions known as the $SL(2,\mathbb{Z})$ family of black holes, first discussed in \cite{Maldacena:1998bw} and elaborated upon in \cite{Dijkgraaf:2000fq}. They give an infinite number of phases, instead of the two we usually consider in Lorentzian signature, and we can check volume independence in each of the phases individually. Twists do not seem to be considered in the literature on large-$N$ volume independence, but we will show that volume independence continues to hold. 

A general $SL(2,\mathbb{Z})$ black hole has a unique contractible cycle, sometimes called an $A$-cycle. The non-contractible cycle (sometimes called a $B$-cycle) is only additively defined, since for any $B$-cycle one can construct another $B$-cycle by winding around the $A$-cycle $n$ times ($n\in \mathbb{Z}$) while going over the original $B$ cycle. The usual convention is to set this winding number to zero. Due to this redundancy the $SL(2,\mathbb{Z})$ family is really an $SL(2,\mathbb{Z})/ \mathbb{Z}$ family specified by which cycle at infinity is contractible in the interior \cite{Dijkgraaf:2000fq}. Here $\mathbb{Z}$ acts as $\tau \rightarrow \tau+n$ for modular parameter $\tau$. This data is given by two relatively prime integers $(c,d)$ with $c\geq 0$.  We also need to include the famous examples $(0,1)$ (thermal AdS$_3$) and $(1,0)$ (BTZ). 

In the rest of this section we will ignore numerical prefactors in the free energy density and will only track the dependence on cycle lengths. Let us consider the simplest cases first, thermal AdS$_3$ and BTZ, both with zero angular potential. This means $\tau$ and $-1/\tau$ are pure imaginary. We have
\begin{align}
\textrm{Thermal AdS}: f(\beta,L) \sim \f{\beta/L}{\beta L} = \f{1}{L^2}\,,\\
\textrm{BTZ}: f(\beta,L) \sim \f{L/\beta}{\beta L} = \f{1}{\beta^2}\,.
\end{align}
These exhibit volume independence for the center-symmetry preserving (i.e. non-contractible) cycles. Let us now add an angular potential $\theta$, which makes $\tau$ complex:
\begin{align}
\textrm{Thermal rotating AdS}&: f(\beta,L;\theta)  \sim \f{\beta/L}{\beta L} = \f{1}{L^2}\,,\\
\textrm{Rotating BTZ}&: f(\beta,L;\theta)  \sim \f{1}{\beta L}\,\f{\beta/L}{\theta^2/L^2+\beta^2/L^2} = \f{1}{\theta^2+\beta^2}\,.
\end{align} 
We again get consistent results, since the lengths of the contractible cycles of thermal rotating AdS and rotating BTZ are $L$ and $\sqrt{\theta^2+\beta^2}$, respectively.

The general $SL(2,\mathbb{Z})$ black hole can be given in a frame where the modular parameter is $(a \tau+b)/(c\tau+d)$, the contractible cycle $z =\phi+it_E \sim z+L(c\tau+d)$ and the non-contractible cycle $z \sim z+L(a\tau+b)$. Their lengths are given as
\be
|S^1_A| = \sqrt{d^2L^2+2 c d L\theta+c^2(\beta^2+\theta^2)}, \qquad |S^1_B| = \sqrt{b^2L^2+2a b L \theta+a^2(\beta^2+\theta^2)}\,.
\ee
The free energy density is found, for general $\tau = i\beta/L+\theta/L$, to be
\be
f(\beta,L;\theta) \sim \frac{{\color{red} a d-b c}}{d^2L^2+2 c d L\theta+c^2(\beta^2+\theta^2)}=\f{1}{|S_A^1|^2}\,.
\ee
Notice that $a$ and $b$ enter into the size of the non-contractible cycle, but the condition $ad-bc=1$ forces the free energy density to be independent of these parameters! This is as expected since the physically distinct states should only care about $c, d$ by the arguments above. We therefore find for the general $SL(2,\mathbb{Z})$ geometry that the free energy density exhibits volume independence.

\subsection{$SL(d,\mathbb{Z})$ family of black holes}
There exists an unexplored analog to the $SL(2,\mathbb{Z})$ family of black holes in higher dimensions, which I will call the $SL(d,\mathbb{Z})$ family of black holes. For a review of some salient points about conformal field theory on $\mathbb{T}^d$ and $SL(d,\mathbb{Z})$, see appendix \ref{app}. 

The bulk topology is that of a solid $d$-torus, with a unique contractible cycle. Winding a $B$-cycle by an $A$-cycle is topologically trivial. A ``small" bulk diffeomorphism, i.e. one continuously connected to the identity, can undo this winding. However, winding a $B$-cycle by another $B$-cycle leads to a true winding number and is topologically distinct. This corresponds to a large diffeomorphism in the bulk. Thus, as in the two-dimensional case, we only need to sum over a subgroup of the full $SL(d,\mathbb{Z})$, because $B$-cycles are only additively defined.  In particular, we consider again the group $SL(d,\mathbb{Z})/\mathbb{Z}$, where $\mathbb{Z}$ acts as $\vec{V}_i\rightarrow \vec{V}_i+n\vec{V}_d$ for all $1\leq i \leq d-1$, $n\in \mathbb{Z}$ and $\vec{V}_d$ the fixed contractible cycle vector. As reviewed in appendix \ref{app}, the $\vec{V}_i$ represent lattice vectors that define the quotient of the plane that gives us the torus $\mathbb{T}^d$.

Our ``seed" solution in three bulk dimensions was global AdS$_3$ at finite temperature and finite angular velocity. In higher dimensions our seed solution will be the AdS soliton, with all spatial directions compactified, arbitrary twists turned on (including both twists between spatial directions and time-space twists, interpreted as angular velocities), and the $d^{\textrm{th}}$ direction contractible. Summing over the restricted set of $SL(d,\mathbb{Z})/\mathbb{Z}$ images of this geometry described above should give an $SL(d,\mathbb{Z})$-invariant partition function. Ignoring the important issue of convergence of this sum, we can see that the invariance is naively guaranteed since the seed solution and its images are independently invariant under the $\mathbb{Z}$ we mod out by.  In other words, the analog of $Z_{0,1}(\tau)$ from the previous section, call it $Z_{0}(\vec{V}_1,\dots \vec{V}_d)$, and its images are invariant under shifts $\vec{V}_i\rightarrow \vec{V}_i+n \vec{V}_d$.  

Anyway, this restricted sum is not important for our purposes. It is sufficient to show that an arbitrary element of the $SL(d,\mathbb{Z})$ family has a free energy density that depends only on the contractible cycle. The simplest case is the AdS soliton at finite temperature with spatial directions compactified, which has free energy density
\be
f(L_1,\dots,L_d) \sim \f{1}{L_d^{d}}\,,
\ee
where $L_d$ is the length of the contractible cycle. This is volume-independent as required. Twisting any of the non-contractible directions by any of the other directions by any amount does not change this answer. Thus, the general AdS soliton with arbitrary angular potentials and spatial twists exhibits volume independence with respect to the non-contractible cycles. We can now consider $SL(d,\mathbb{Z})$ images of this geometry.

The general $SL(d,\mathbb{Z})$ image geometry has global Killing vector fields for all the non-radial coordinates, which reduces the on-shell action calculation to $\int d^{d+1}x \sqrt{g} \,F(r, \tilde{r}_h)=\textrm{Vol}(\mathbb{T}^{d})\int dr F(r, \tilde{r}_h)$ where $\tilde{r}_h$ is a parameter fixed by the size of the $d^{\textrm{th}}$ cycle and $F(r,\tilde{r}_h)$ is some function. Thus, twists can only enter into $\textrm{Vol}(\mathbb{T}^{d})$, but torus volumes are invariant under twists. Higher-order corrections in the Newton constant $G_N$ will bring in a dependence on the twists, as the momentum quantization of perturbative fields on a torus depends on the twists. In this way we see that volume-independence will break down  at subleading order in the $1/N$ expansion, as expected. Let us calculate the free energy density a little more carefully.

Consider a general twisted seed geometry, with the contractible direction chosen to lie along the $d^{\textrm{th}}$ direction, specified by lattice vectors defining the twists:
\begin{equation}
\Theta=\begin{bmatrix}
\theta_{11}  & \theta_{12}   & \cdots   &\theta_{1,(d-1)}& \theta_{1d}  \\
0  & \theta_{22}    & \cdots &\theta_{2,(d-1)}& \theta_{2d} \\
\vdots  & \vdots & \ddots   & \vdots& \vdots\\
0  &  0 &\cdots     &\theta_{(d-1), (d-1)} & \theta_{(d-1),d}  \\
0  &  0 & \cdots     &0&\theta_{dd}
\end{bmatrix}\,.
\end{equation}
This can be transformed by a general $SL(d,\mathbb{Z})$ transformation $(A)_{ij} = a_{ij}\in \mathbb{Z}$ with $\det(A)=+1$ to give
\begin{equation}
A\cdot \Theta =\begin{bmatrix}
a_{11}\theta_{11}  & a_{11}\theta_{12}+a_{12}\theta_{22}   & \cdots   & \sum_{i=1}^d a_{1i}\theta_{id}  \\
a_{21}\theta_{11}  & a_{21}\theta_{12}+a_{22}\theta_{22}   & \cdots  & \sum_{i=1}^d a_{2i}\theta_{id} \\
\vdots  & \vdots & \ddots   & \vdots  \\
a_{d1}\theta_{11}   &  a_{d1}\theta_{12}+a_{d2}\theta_{22} & \cdots  & \sum_{i=1}^d a_{di}\theta_{id}
\end{bmatrix}\,.
\end{equation}
We can compose $d(d-1)/2$ rotations in the $d(d-1)/2$ two-planes to make this matrix upper triangular. This will allow us to identify the new modular parameter matrix $\Theta_{\textrm{new}}$. This will not change the lengths of the cycles, which are given as 
\be
L_k^2 = \sum_{j=1}^d \left(\sum_{i=1}^j a_{ki}\theta_{ij}\right)^2\,,
\ee
where $L_d$ gives the length of the contractible direction. The volume of the resulting torus is given as 
\be
\text{Vol}(A \mathbb{T}^d) = \det(A \cdot \Theta) = \det(A)\det(\Theta) = \prod_{i=1}^d \theta_{ii}\,.
\ee
In particular, it is unchanged by the $SL(d,\mathbb{Z})$ transformation. The free energy density is given as
\be
f(L_1,\dots,L_d) \sim \f{1}{L_d^d}\,,
\ee
exhibiting volume independence in the center-symmetric directions.

Notice that, like for the case $d=2$, twists in the contractible direction are redundant with the case of no twist in that direction. This is because there exists a bulk diffeomorphism, continuously connected to the identity, which induces this twist on the boundary. Twists in non-contractible directions, however, correspond to large gauge transformations and define distinct geometries. So, unlike for $d=2$, uniquely defining the contractible cycle is not sufficient. We still have a reduction in moduli, with $d^2-1-(d-1) = d(d-1)$ numbers specifying distinct geometries. Interestingly, the distinct geometries obtained by twisting non-contractible directions by other non-contractible directions do not differ in their classical on-shell action. 

\section{Correlation functions and entanglement entropy}\label{corr}
In this section we will discuss the implications of the Eguchi-Kawai mechanism for correlation functions and Renyi entropies. As usual, the statements are restricted to leading order in $N$, meaning tree-level Witten diagrams in the bulk. We will only consider volume independence with respect to a single direction for conceptual clarity; generalization to multiple directions is straightforward. For correlation functions we will see that position space correlators must be given by the method-of-images under smooth quotients, as in \eqref{volindpos}. The connection between large-$N$ reduced correlation functions and the role of the method of images in AdS has previously been explored in the stringy (zero 't Hooft coupling) limit in \cite{Furuuchi:2005qm, Furuuchi:2005eu}, although there are several points of deviation from the present work.
\subsection{Correlation functions}
Let us assume that we are volume-independent with respect to a single direction. Then connected correlation functions of local, single-trace, gauge-invariant, neutral-sector observables will be volume independent at leading order in $N$. Nonlocal operators like Wilson loops can also be treated as long as they have trivial winding around the cycle. One term that may need explanation is ``neutral-sector." We will explain briefly below; for details see \cite{Kovtun:2007py}.

Consider the theory on $\mathbb{R}^{d-1}\times S^1$ as we vary the circle size from some length $L$ to some other length $L'$. A given operator in the theory of size $L$ can be decomposed as
\be
O(x) = \sum_{n=-\infty}^{\infty} O_{n/L} e^{2\pi i n x/L}\,.
\ee
$O_{n/L}$ for momenta $n/L$ commensurate with those of the box of size $L'$ are termed ``neutral-sector" operators, and it is their correlation functions which are volume-independent. For example, if $L=x L'$ for some irrational $x$ then only correlators of $O_0$ will be volume-independent. While this may seem like a severe restriction, we will only be concerned with producing finite-size results from infinite-size results, and all momenta in finite size are commensurate with some momentum in infinite size. 

For a theory of pure glue or pure adjoint fields like $\mathcal{N}=4$ super Yang-Mills, we can write this precisely as 
\be
\lim_{N\rightarrow \infty} N^{2(M-1)}\langle O_{n_1/L}O_{n_2/L}\dots O_{n_M/L}\rangle_L= \lim_{N\rightarrow \infty} (JN^2)^{M-1}\langle O_{n_1/L}O_{n_2/L}\dots O_{n_M/L}\rangle_{JL}\,
\ee
 for $J\in \mathbb{Z}^+$ and $L'=JL$. This particular limit is because the large-$N$ counting in a purely adjoint theory shows that the connected correlator of $M$ single-trace operators is $\mathcal{O}(1/N^{2(M-1)})$. For a theory with mesonic operators and therefore an expansion in $1/N$, with connected correlators of mesons scaling as $\mathcal{O}(1/N^{M-1})$, we would take a limit with $N^{M-1}$ in front to isolate the leading contribution to the connected correlator. But the basic point is clear: the statement is about the first order in $N$ that is expected to have a nonvanishing answer by large-$N$ counting. If it vanishes, no statements are made about the leading nonvanishing order. This is what the limit above makes precise in a pure adjoint theory. We will not worry about the various cases of large-$N$ counting, because within AdS/CFT the leading-in-$N$ diagrams are given by tree-level diagrams in the bulk. It is only these diagrams we wish to make a statement about. We will therefore use as our primary tool the equality
\be\label{hammer}
\langle O_{n_1/L}O_{n_2/L}\dots O_{n_M/L}\rangle_L=  J^{M-1}\langle O_{n_1/L}O_{n_2/L}\dots O_{n_M/L}\rangle_{JL}\,
\ee
with the caveat that this is the leading-in-$N$ piece of a connected correlator left implicit.

To see the effect on a general correlation function of local operators, it will suffice to consider the two-point function. We consider the Fourier representation of the finite-size correlator:
\begin{align}
\langle O(x)O(y)\rangle_L &= \sum_{(n,m) \in \mathbb{Z}^2} e^{-2\pi i(nx+my)/L}\langle O_{n/L}O_{m/L}\rangle_L\,\\
&=\sum_{(n,m) \in \mathbb{Z}^2} e^{-2\pi i(nx+my)/L}J\langle O_{n/L}O_{m/L}\rangle_{JL}\,,
\end{align}
where in the second line we used \eqref{hammer}. We could immediately use translation invariance to write the correlator as a function of only the separation $x-y$, but to make generalization to higher-point correlators clear we will keep the dependence until the end. 

We can now simplify this expression by transforming the momentum-space correlator in size $JL$ to position space and evaluating the various sums and integrals:
\begin{align}
\langle O(x)O(y)\rangle_L 
&=\hspace{-1mm}\sum_{(n,m) \in \mathbb{Z}^2}\f{1}{J^2L^2}\int_0^{JL}\hspace{-1mm}dx'\int_0^{JL}\hspace{-1mm}dy' e^{-2\pi i(nx+my)/L}e^{2\pi i(nx'+my')/L}J\langle O(x')O(y')\rangle_{JL}\\
&=\f{1}{JL^2}\int_0^{JL}\hspace{-1mm}dx'\int_0^{JL}\hspace{-1mm}dy' \sum_{(n,m) \in \mathbb{Z}^2}\hspace{-2mm}e^{-2\pi i(nx+my)/L}e^{2\pi i(nx'+my')/L}\langle O(x')O(y')\rangle_{JL}\\
&=\f{1}{JL^2}\int_0^{JL}\hspace{-1mm}dx'\int_0^{JL}\hspace{-1mm}dy'\hspace{-2mm} \sum_{(n,m) \in \mathbb{Z}^2}\hspace{-2mm}L^2 \delta(x'-x-nL)\delta(y'-y-mL)\langle O(x')O(y') \rangle_{JL}\\
&=\f{1}{J} \sum_{n,m=0}^{J-1}\langle O(x+nL)O(y+mL)\rangle_{JL}\,.
\end{align}
This generalizes to
\be\label{volindpos}
\langle O(x_1)\dots O(x_M)\rangle_L = \f{1}{J} \sum_{n_i=0}^{J-1} \langle O(x_1+n_1L)\dots O(x_M+n_M L)\rangle_{JL}\,.
\ee
The converse is also true. That is, starting from the method-of-images form of a position space correlator above, one can show \eqref{hammer}. Altogether, volume-independence of neutral-sector correlators is true if and only if finite-size correlators are obtained by the method of images from correlators in a larger size.


\subsection{Two-point functions}
To focus on the simplest case, consider the equal-time two-point function in a translation-invariant two-dimensional theory. Say we want to construct the finite-size correlator from the infinite-size correlator. We begin from \eqref{volindpos} and use translation invariance, which says that our correlator is only a function of the distance between the two insertion points:
\begin{align}
\langle O(x)O(y)\rangle_L &=\langle O(x-y)\,O(0)\rangle_L =\f{1}{J} \sum_{n,m=0}^{J-1} \langle O(x-y+(n-m)L)\,O(0)\rangle_{JL}\\
&=\f{1}{J} \sum_{n,m=0}^{J-1} \langle O(x-y+(n+(J-m))L)\,O(0)\rangle_{JL}\\
&=\sum_{n=0}^{J-1}\langle O(x-y+nL)\,O(0)\rangle_{JL}\,,
\end{align}
where we used the $JL$-periodicity of the size-$JL$ correlator. To compare to the infinite-size correlator we can take $J\rightarrow \infty$ in a particular way:
\begin{align}
\langle O(x-y)\,O(0)\rangle_L &= \lim_{J\rightarrow \infty}\f{1}{2} \sum_{n=0}^{J-1}\left[ \langle O(x-y+nL)\,O(0)\rangle_{JL}+\langle O(x-y+(n-J)L)\,O(0)\rangle_{JL}\right]\\
&=\lim_{J\rightarrow \infty}\f{1}{2}\sum_{n=-(J-1)}^{J-1} \langle O(x-y+nL)\,O(0)\rangle_{JL} \,.
\end{align}
Notice that taking this limit will give us the correlator on the semi-infinite line with semi-infinite periodicity. Doubling it (and picking up a factor of $2$ just as in the factor of $J$ that comes from relating two-point functions in size $L$ to size $JL$) gives us the real-line correlator. We thus have our final result
\be
\langle O(x-y)\,O(0)\rangle_L =\sum_{n=-\infty}^\infty \langle O(x-y+nL)\,O(0)\rangle_\infty\,.
\ee

Now we compare to gravity in AdS. Conformal field theory correlators, at leading order in $N$, are obtained by extrapolating the bulk-to-bulk propagator to the boundary. Since the bulk-to-bulk propagator for free fields satisfies a Green function equation, we can find the propagator after performing an arbitrary smooth quotient by the method of images. This gives precisely the form of correlator above, which for example in the famous case of the BTZ black hole takes the form \cite{KeskiVakkuri:1998nw}
\be
\langle O(t,\phi)O(0,0)\rangle = \sum_{n=-\infty}^\infty\f{1}{\left(\cosh\left(\f{2\pi t}{\beta}\right)-\cosh\left(\f{2\pi (\phi+n L)}{\beta}\right)\right)^{2\Delta}}
\ee
for operators of dimension $\Delta$. Notice that this sums over spatial images but not thermal images. For thermal AdS$_3$, which is obtained instead as a quotient in the Euclidean time direction, we would sum over thermal images but not over spatial images. In each case, the correlator is given by a sum over images with respect to the center-preserving direction. This is exactly what is predicted by our arguments above. Furthermore, we see that the ``free-ness" of large-$N$ theories is not sufficient by itself to imply that the correlator should be a sum over images, since there is no sum over images in the center-breaking direction.


\subsection{$M$-point functions}
\begin{figure}
\centering
\includegraphics[scale=0.32]{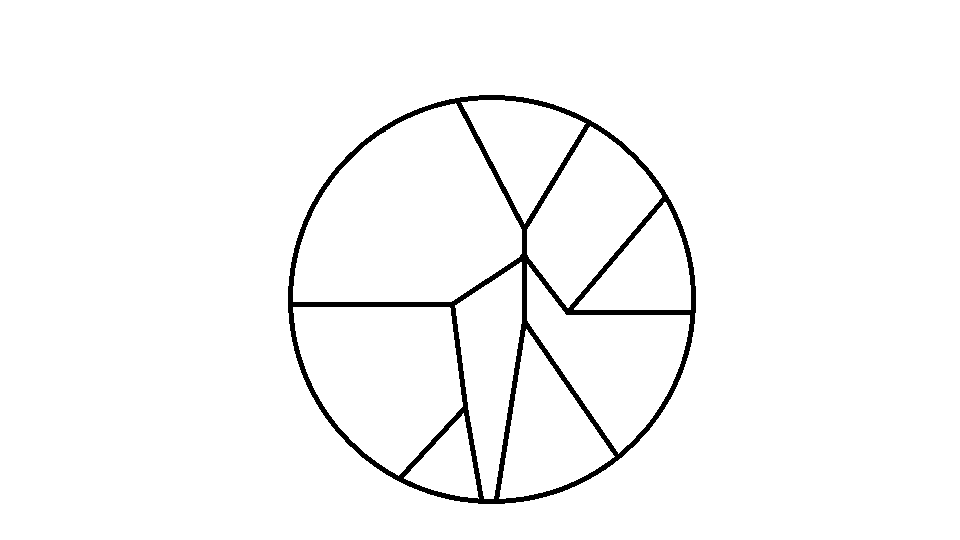} \includegraphics[scale=0.32]{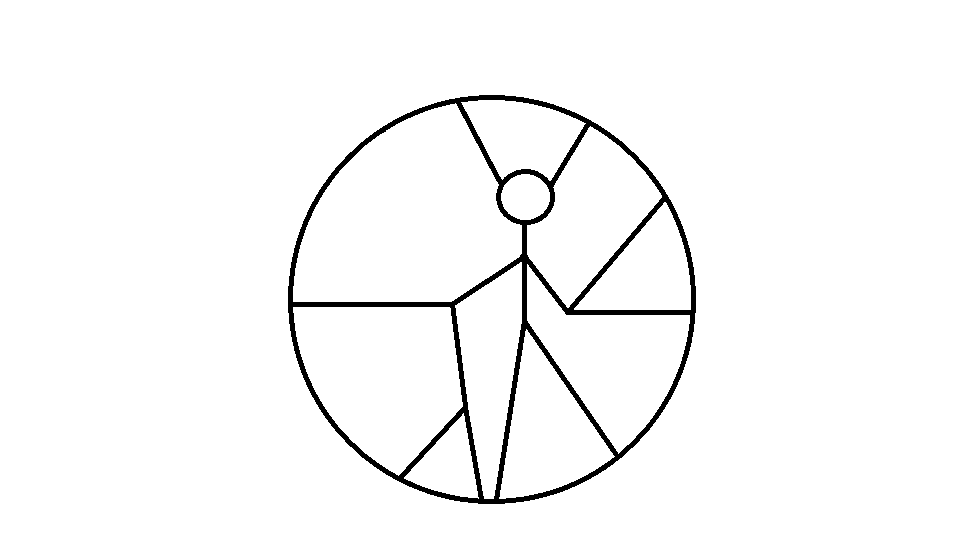}
\caption{\label{wittendiag} Left: A tree-level Witten diagram, which contributes at leading order in $N$ to the nine-point function. It is constructed out of $M=9$ bulk-to-boundary propagators and $n=5$ bulk-to-bulk propagators. Since it is a contribution at tree level, there are $6=n+1$ interaction vertices. There are many more diagrams contributing at this order. Right: A loop-level Witten diagram, which contributes at first subleading order in $N$ to the nine-point function. It is constructed out of $M=9$ bulk-to-boundary propagators and $n=8$ bulk-to-bulk propagators. There are $8\neq n+1$ interaction vertices. There are again many more diagrams contributing at this order.}
\end{figure}

For higher-point functions, recall that we focus only on diagrams in the bulk that do not have any loops. Any given contribution to the tree-level $M$-point function is constructed out of $M$ bulk-to-boundary propagators $K$ and $n<M$ bulk-to-bulk propagators $G$. This means there are $n+1$ interaction vertices in the bulk. An illustrative case of tree-level (leading in $N$) and loop level (subleading in $N$) diagrams is depicted in figure \ref{wittendiag}.

The position space correlation function can be written schematically as 
\be
\langle O(x_1)\dots O(x_M)\rangle_{AdS} = \int_{AdS} \prod_{i=1}^{n+1} dX_i \,G^n\, K^M\,,
\ee
where boundary points are denoted by small $x$ and bulk points by big $X$. From here we can show
\be\label{toprove}
\langle O_{n_1/L}\cdots O_{n_M/L}\rangle_{AdS/\Gamma}=|\Gamma|^{M-1}\langle O_{n_1/L}\cdots O_{n_M/L}\rangle_{AdS}\,.
\ee
Before we outline the proof of this we need the following facts. The bulk-to-bulk propagator satisfies a Green function equation since the bulk theory is free at this order (leading in $N$). The bulk-to-boundary propagator is obtained by a certain limit of the bulk-to-bulk propagator where one of its points is pulled to the boundary. Thus, both propagators can be obtained on a smooth quotient of our AdS background by the method of images. Finally, in momentum space, the integrals over spacetime give $n+1$ momentum-conserving delta functions since there are no loops in the bulk.

The general proof of \eqref{toprove} is notationally clumsy and would ruin the already regretful aesthetics of this paper, so we will provide an outline of the general proof here and give a sample calculation in appendix \ref{fourpoint}. The left-hand-side is evaluated by an inverse Fourier transform of the position space expression. The position space expression is written in terms of the bulk-to-bulk and bulk-to-boundary propagators in AdS$/\Gamma$. These are replaced by those in AdS by the method of images. These propagators are then transformed into momentum space (with momenta now spaced according to AdS and not AdS$/\Gamma$). Sums and integrals are re-ordered at will and this expression is simplified down to an integral over the bulk radial interaction vertices $z_i$. The right-hand-side is evaluated in the same way, except its propagators are never replaced with other propagators. This leads to \eqref{toprove}. Explicit details for a four-point function can be found in  appendix \ref{fourpoint}. 

So we see that the behavior of tree-level perturbation theory in AdS$_{d+1}$ under generic, smooth quotients of spacetime is reproduced. Notice that bulk loops are made of bulk-to-bulk propagators as well, but their momenta are not fixed and instead are integrated over. This leads to a non-universal answer, since there are bulk-to-bulk propagators in the AdS space and AdS$/\Gamma$ space which have explicit momentum dependence which is to be summed over. Although the bulk-to-bulk propagator in AdS$/\Gamma$ can be replaced with the one in AdS by the usual method of images trick, the sums are over different momenta and cannot be carried out in general.

\subsection{Entanglement/Renyi entropies}
Another place where volume independence crops up is in the calculation of entanglement entropy of theories dual to gravity in AdS$_{d+1}$. For simplicity I will restrict to AdS$_3$.

Recall that the Ryu-Takayanagi prescription dictates that the entanglement entropy is given by the regularized area of a minimal surface that is anchored on the entangling surface on the AdS boundary \cite{Ryu:2006ef}. Consider a spatial interval of size $\ell$ on a spatial circle of size $L$ at temperature $T$. For entangling surfaces at fixed time for static states or density matrices, the minimal surface will lie on a constant bulk time slice. This makes it clear that in the confined phase, which is thermal AdS$_3$, the Ryu-Takayanagi answer will be independent of the center-preserving thermal circle of size $T$:
\be
S_{EE} = \f{c}{3} \log \left(\f{L}{\epsilon} \sin \left(\f{\pi\ell }{L}\right)\right)
\ee
for UV cutoff $\epsilon$. Note that it is not given as a sum over thermal images like in the case of correlation functions. It is instead completely independent of the thermal cycle size.

In the deconfined phase, i.e. above the Hawking-Page CSST, we get an answer independent of the center-preserving spatial circle of size $L$:
\be
S_{EE} = \f{c}{3} \log \left(\f{\beta}{\epsilon} \sinh \left(\f{\pi\ell }{\beta}\right)\right).
 \ee 
The minimization inherent in the Ryu-Takayanagi prescription is the reason why we do not sum over images and so get exact volume independence. (There is a proposal that the image minimal surfaces instead contribute to entanglement between internal degrees of freedom, coined ``entwinement" \cite{Balasubramanian:2014sra}.) 

Apparently, single-interval entanglement entropy is an appropriate neutral-sector ``observable" that obeys large-$N$ volume independence. As shown by a bulk calculation in \cite{Barrella:2013wja}, volume-dependence appears at first subleading order in the central charge $c$ (the proxy for $N$ in two-dimensional theories). Volume-dependence also appears at leading order in the central charge in the Renyi entropies, but not in any trivial way as in the local correlators of the previous section. The Renyi entropies must not be neutral-sector observables. The Renyi entropy in this context is related to the free energy on higher-genus handlebodies; the analytic continuation connecting to the original torus to define the entanglement entropy is therefore special.

It is interesting that in the cases where we have a volume-independent object, it is the entanglement entropy and not any of the higher Renyi entropies. This may be related to the fact that it is the entanglement entropy that naturally geometrizes in the bulk, or to the fact that it is a good ensemble observable (or these two could be the same thing). 

\section{Higher-spin theory as a failure of the Eguchi-Kawai mechanism}\label{higherspin}
We have presented large-$N$ volume independence along all but one cycle of toroidal compactifications as a necessary condition for a field theory to have a local gravitational dual. This is discussed further in section \ref{necek}. Higher-spin theories are a good example of how things go wrong if this does not occur, and provide additional evidence for this conjecture. Higher-spin theories in AdS are nonlocal on the scale of the AdS curvature.

There are a zoo of higher-spin theories, so let us analyze one of the simplest cases. Consider the parity-invariant Type-A non-minimal Vasiliev theory with Neumann boundary conditions for the bulk scalar field \cite{Fradkin:1986qy, Fradkin:1987ks, Vasiliev:1999ba}. This is a theory that can be expanded around an AdS$_4$ background and has fields of all non-negative integer spin. It is proposed to be dual to the three-dimensional, free $U(N)$ vector model of a scalar field restricted to the singlet sector \cite{Klebanov:2002ja}. The singlet projection is performed by weakly gauging the $U(N)$ symmetry with a Chern-Simons gauge field.

The Chern-Simons-matter theory does not enjoy large-$N$ volume independence. In fact, given that the matter is in the fundamental representation, it does not even have center symmetry at the Lagrangian level. However, there is a simple procedure for deforming such theories into close cousins with explicit center symmetry at the Lagrangian level. This is discussed for example in \cite{Kovtun:2007py}. First we add a global $U(N_f)$ flavor symmetry to the boundary theory, and then we weakly gauge it and change the representation of the matter to be in the bifundamental. Such a theory has explicit center symmetry at the Lagrangian level and will have center-symmetric phases (as long as e.g. $N=N_f+\mathcal{O}(N_f/N)$). There now exist single-trace, gauge-invariant operators made up of arbitrarily long strings of the bifundamental fields, which did not exist in the previous theory. These are the objects associated to the string states in the bulk. 

This procedure, with some more bells and whistles (the bells and whistles being an appropriate amount of supersymmetry), is precisely what takes these vector models into the more mature ABJ theory \cite{Aharony:2008ug, Aharony:2008gk}. The bulk interpretation of this procedure is also straightforward and deforms the higher-spin theory into its more mature cousin, string theory. The addition of the global flavor symmetry is the addition of Chan-Paton factors to the higher-spin theory, which implies upgrading the spin-$1$ bulk gauge field to a non-abelian $U(N_f)$ gauge field, with all other fields transforming in the adjoint of $U(N_f)$. The gauging is then a familiar procedure in AdS/CFT whereby the boundary conditions of this bulk gauge field are changed. In fact, this entire story is just that of the ABJ triality beautifully painted in \cite{Chang:2012kt}, whereby the higher-spin ``bits" are conjectured to bind together into the strings of ABJ theory. All I would like to highlight is that the deformations that were necessary to connect to a theory with a local gravitational limit included deforming to a theory with an explicitly center-symmetric Lagrangian and center-symmetric phases (at leading order in $N$ and the 't Hooft coupling $N/k$). Interestingly, this deformation also leads to a lifting \cite{Banerjee:2013nca} of the light states present in vector models \cite{Banerjee:2012gh}.

It may be interesting to explore what other deformations of the vector models can introduce center symmetry and the particular center symmetry structure that is a hallmark of classical gravity. This may shed light on how to deform the set of proposed higher-spin dualities for de Sitter space \cite{Anninos:2011ui, Chang:2013afa, Anninos:2014hia} to an Einstein-like dual. In the context of de Sitter, the deformation discussed above leads to a ``tachyonic catastrophe" in the bulk, as discussed in \cite{Anninos:2013rza}, and does not seem to give a viable option.

\section{Learning about the Eguchi-Kawai mechanism from gravity}\label{fieldtheory}
In this section, we will shift our focus and analyze what gravity teaches us about the Eguchi-Kawai mechanism.
\subsection{Center symmetry stabilization and translation symmetry breaking}\label{percen}
Although this was discussed in previous sections, we would like to emphasize that the bulk gravitational description gives us a way to predict whether volume independence is upheld in particular holographic gauge theories. The first nontrivial statement is that center symmetry can be broken along at most one cycle for any given configuration of cycle sizes. The second nontrivial statement is that there are simple ways to preserve center symmetry along a given cycle for any cycle size which remains larger than string scale in the bulk. In particular, periodic bulk fermions and antiperiodic bulk scalars prevent cycles from capping off in the bulk, as this is an inconsistent spin structure. These cases therefore preserve center symmetry beyond the CSST points which correspond to gravitational Hawking-Page transitions. This argument does not explicitly rely on the representation of the boundary matter (for example it is true of $\mathcal{N}=4$ super Yang-Mills and for ABJ theory, where the fermions are in the adjoint and the bifundamental, respectively). The bulk matter is made of gauge-invariant combinations of the boundary fields, so the periodicity conditions of the bulk matter will be correlated with the periodicity conditions of the boundary fields. For example, bulk fermions are constructed by taking single-trace gauge-invariant operators consisting of an odd number of boundary fermionic fields (e.g. $\tr[\phi\psi]$). Therefore, bulk fermions with periodic spin structure imply boundary fermions with periodic spin structure. A similar statement is true for antiperiodic bulk scalars. Higher-spin dualities, however, offer an interesting case where the bulk theory is purely bosonic while the boundary theory can be purely fermionic. 

The quantum-mechanically generated potentials for the gauge field holonomies can be straightforwardly calculated at weak coupling, see for example \cite{Gross:1980br, Shifman:2008ja}. From the weakly coupled point of view, for $(3+1)$-dimensional $SU(N)$ Yang-Mills theories, preserving center symmetry with non-adjoint periodic fermions or antiperiodic scalars of any representation is not possible. The only choice that works is periodic adjoint fermions. Interestingly, for periodic adjoint fermions (which we will have for super Yang-Mills theories) we seem to preserve center symmetry at strong coupling as well. But there is a small catch. At weak coupling, one would need to make the fermions periodic along all $k$ cycles of $\mathbb{T}^k \times \mathbb{R}^{d-k}$.\footnote{A calculation of the Casimir energy in $\mathcal{N}=4$ super Yang-Mills on $\mathbb{T}^2 \times \mathbb{R}^2$ \cite{Myers:1999psa}, for example, shows that we lose volume independence along both cycles if the fermion is periodic along only one cycle.} At strong coupling, however, this will not give us a background well-described by gravity alone, since it will be the toroidally compactified Poincar\'e patch with circles shrinking to substringy scales near the horizon. To have a proper gravitational description, we would need to make the fermions antiperiodic along one of the cycles (or the scalars periodic). In this case, we will still preserve center symmetry along all the cycles that have periodic fermions, but this does not match what happens at weak coupling. 

We have not made any comments about operator expectation values and correlation functions within a grand canonical ensemble, say for turning on a chemical potential for some global symmetry. In this case, one can spontaneously break translation invariance, in which case the Eguchi-Kawai mechanism fails \cite{Unsal:2010qh}. There exist holographic examples of such spatially modulated phases  \cite{Nakamura:2009tf, Donos:2011bh, Anninos:2013mfa}.


\subsection{Extending the Eguchi-Kawai mechanism to curved backgrounds}
An important question about the Eguchi-Kawai mechanism is whether it extends to curved backgrounds. The original Eguchi-Kawai mechanism, and most modern proofs of large-$N$ volume independence, rely on a lattice regularization which we do not have on curved backgrounds (although see \cite{Brower:2012vg} for some progress in the case of spherical backgrounds). We will set this aside for the moment as a technical issue. We will see that the natural uplift of volume independence to curved backgrounds is what I will call ``topological volume independence." We will make this notion precise by defining an order parameter (which will again be the expectation value of a Polyakov loop) and checking in gravitational examples that ``topological volume independence" is indeed realized.

\subsubsection*{Hints from field theory}
We already have some hints from field theory about what the Eguchi-Kawai mechanism on curved manifolds should look like. The first hint comes from the perturbative intuition for volume independence on torus compactifications. In particular, mesons and glueballs form the confined phase degrees of freedom (baryons have masses that scale with $N$ and can be ignored for our purposes), and interactions between these confined phase degrees of freedom are suppressed by $1/N$. At leading order in $N$ the theory behaves as if it is free. The confined phase degrees of freedom are therefore incapable of communicating with their images to discover they are in a toroidal box. This intuition, however, is valid even in a curved box. This seems to suggest the size of the manifold should again not be relevant even if it is curved. But curved backgrounds have local curvature which can vary as you change the overall size of the background, e.g. increasing the radius of a sphere. There is no reason the mesons and glueballs cannot feel this local curvature at leading order in $N$ and thereby (for maximally symmetric manifolds like a sphere of hyperboloid) would know the overall size of the compact manifold on which they live. So it seems we should not expect a totally general uplift to curved backgrounds.

The second hint comes from thinking about volume independence in toroidal compactifications as a generalized orbifold projection, where one orbifolds by a discrete translation group \cite{Kovtun:2007py}. (The language of orbifolds here is conventional but everything is really a smooth quotient.) Generic changes in the overall size of curved backgrounds cannot be thought of this way, so we again see that we cannot expect a totally general uplift to curved background.

Combining the two hints above provides a compelling case for what kind of setup has a chance of maintaining a useful notion of volume independence. One begins with a curved background and considers smooth quotients that change the volume of the manifold. Such operations do not change the local curvature and maintain the picture of volume-changing as an orbifold procedure. This therefore utilizes the two hints above. We can now check that gravity provides a calculable setup where this proposal for the Eguchi-Kawai mechanism on curved backgrounds can be checked to be valid. The simplest case to analyze is the conformal field theory on any simply connected manifold, like the sphere or the hyperboloid. As an illustrative example, we will investigate the family of lens spaces formed by smooth quotients of $S^3$, although our results are general. For any smooth quotients of simply connected manifolds, we will see that the Polyakov loop expectation value continues to serve as an order parameter for center symmetry.

\subsubsection*{Holographic realization of the Eguchi-Kawai mechanism on curved manifolds}
Holographic gauge theories in the gravitational limit realize all of the intuition of the above. They explicitly show that naive volume-independence on curved backgrounds does not hold. Furthermore, they show that topological volume independence \emph{does} hold when interpreted in the above sense! 

To see that naive volume independence on curved backgrounds does not hold, we can consider an observable as basic as the zero-point function, or the free energy density. We saw that for torus-compactified holographic theories, the free energy density was volume-independent due to the thermodynamics of black branes. For holographic theories on a sphere or the hyperboloid, this is no longer the case. The relevant bulk geometries are the spherical and hyperbolic black holes. The key difference between these geometries and the black brane is that the horizon radius is not proportional to the Hawking temperature. Instead, we have
\be
T \sim r_h + r_h^{-1}\,.
\ee
This means that the Bekenstein-Hawking area law, which scales as $r_h^{d-1}$ and gives the thermal entropy of the CFT, is not extensive in field theory variables (i.e. does not scale as $T^{d-1}$). Here it is important to keep in mind that the theories we are considering are conformal,  so fixing the temperature dependence fixes the volume dependence. Moreover changing the radius of the sphere or hyperboloid can equally well be regarded as changing the temperature since only the ratio is meaningful. But since $S_{CFT} = S_{BH} = r_h^{d-1} \neq T^{d-1}$, we do not have extensivity of the thermal entropy or the free energy, unless $r_h\rightarrow \infty$ which pushes us into the black brane limit. Furthermore, correlation functions in these backgrounds have nontrivial volume-dependence. While the ideas of large-$N$ volume independence do not apply, there may still be a lower-dimensional matrix model description of the higher-dimensional theory, see e.g. \cite{Ishii:2008ib, Kawai:2009vb, Honda:2012ni}

Both of these problems are solved by considering the smooth orbifolds suggested in the previous section. The entropy density (or free energy density) becomes appropriately volume-independent because smooth orbifolds of the spatial manifold cannot be interpreted as changes in the temperature. Thus, the nonlinear relation between horizon radius and temperature is not a problem. Said another way, we consider a setup where our field theory is on a manifold $\mathcal{M}^{d-1}$ and its thermal ensemble at high temperature (i.e. the deconfined theory) is dominated by a black object with horizon topology $\mathcal{M}^{d-1}$. This is what happens for the field theory on a sphere, plane, or hyperboloid. The quotient of the manifold $\mathcal{M}_{d-1}$ by some freely acting group $\Gamma$ changes the Bekenstein-Hawking entropy as follows:
\be
S_{\textrm{BH}} = \f{r_h^{d-1} \,\mathrm{Vol}(\mathcal{M}_{d-1})}{4G_N}\longrightarrow \f{r_h^{d-1} \,\mathrm{Vol}(\mathcal{M}_{d-1}/\Gamma)}{4G_N}\,.
\ee
We see from this formula that the field theory's entropy density and free energy density is appropriately independent of such changes in volume, as long as no CSST occurs (more on this possibility below).

How about correlation functions? As we saw before, these are constructed by bulk Witten diagrams, whose atoms are bulk-to-bulk and bulk-to-boundary propagators. These objects again obey a Green function equation in the bulk, meaning any orbifold of the background geometry can be dealt with by summing over orbifold images. As long as we remain at leading order in $N$, meaning we do not consider bulk loops, the correlator will pick up a trivial volume dependence fully determined by the volume-dependence before quotienting.

We have analyzed volume independence in the deconfined phase of the theory, where the relevant bulk geometries which dominate the thermal ensemble are given by black holes with some horizon topology. Uplifting the intuition from our torus-compactified theories, we should expect to find nontrivial volume-dependence and temperature-independence in the confined phase of the theory. We will address this in the next section.

It is interesting that the gravitational description and the field theory description give the same hints as to what sort of generalization to curved backgrounds should work. In particular, we discussed how from the field theory point of view we should expect volume-changing orbifolds to be the natural uplift of the Eguchi-Kawai mechanism to curved backgrounds. Gravity gives the exact same intuition, and furthermore it explicitly demonstrates that it \emph{works}, at least for the types of observables considered above.

\subsubsection*{Order parameter on curved manifolds and testing topological volume independence}
For any simply connected manifold $\mathcal{M}^{d-1}$, the quotient by some freely acting group $\Gamma$ gives a manifold with nontrivial fundamental group isomorphic to $\Gamma$. This means that we can wrap a Polyakov loop on the existing nontrivial cycle and could reasonably expect that its expectation value continues to serve as a good order parameter. We will see in a concrete example that this is the case. 

To illustrate the point, consider the family of lens spaces $L(p,1)$ which have $\pi_1(L(p,1))=\mathbb{Z}_p$.  These can be understood in terms of the canonical metric for the Hopf fibration:
\be
d\Omega_{3/p}^2=\f{1}{4}\left(d\psi^2+d\theta^2+d\phi^2+2 \cos \theta\, d\psi d\phi\right),\qquad 0<\theta < \pi, \quad 0\leq \phi < 2\pi, \quad 0\leq \psi < 4\pi/p\,.
\ee
Volume independence for lens spaces can now be stated in terms very close to that of the generalized orbifold projections used to discuss volume independence for torus compactifications. Just as we vary the size of a circle in a torus compactification by shifting its periodicity, in this case we move between lens spaces by changing the periodicity of the $\psi$ coordinate. To maintain a smooth quotient we need $p\in \mathbb{Z}^+$ so these are discrete changes. The area of a unit-radius lens space is simply $2\pi^2/p$, and its change is given immediately by the change in the $\psi$ circle.  We can wrap a  Polyakov loop around the $\psi$ circle due to the nontrivial homotopy, and it is again the expectation value of this loop which we propose serves as our order parameter.   

Let us turn to the gravity picture. In the deconfined phase, the orbifolded circle is non-contractible in the bulk, which implies a vanishing Polyakov loop expectation value and therefore volume independence:
\be
ds^2=-\left(1+\f{r^2}{\ell^2}-\f{\mu}{r^2}\right)dt^2+\f{dr^2}{\left(1+\f{r^2}{\ell^2}-\f{\mu}{r^2}\right)}+r^2d\Omega_{3/p}^2\,.
\ee
 As we showed in the previous section, topological volume independence is indeed realized in the free energy density and correlation functions. How about the confined phase? The naive geometry for the confined phase is obtained by taking a quotient of global AdS. This geometry has a conical singularity at the origin which is not well-described within gravity. For antiperiodic fermions along the orbifolded circle (with even $p>2$), it has been proposed that closed string tachyon condensation regularizes the geometry into what is called the Eguchi-Hanson-AdS soliton \cite{Clarkson:2005qx, Clarkson:2006zk}. This geometry has the orbifolded circle smoothly capping off in the interior, giving a nonvanishing expectation value to the Polyakov loop. There is a deconfining CSST at inverse temperature 
\be
\beta_c=2 \pi  \,\frac{\sqrt{-6 p^4+48 p^2-88+\left(p^4-8 p^2+20\right)^{3/2}}}{p^4-8 p^2+4}\,.
\ee
(This corrects the expression given in (4.14) of \cite{Hikida:2006qb}.) In the confined phase, an analysis of the Eguchi-Hanson soliton shows that we have topological volume-dependence with respect to the spatial manifold and volume-independence with respect to the thermal circle! This picture of topological volume independence is also found in ABJM theory through a nontrivial calculation utilizing supersymmetric localization on lens spaces \cite{Alday:2012au}.

An intuitive way to understand the absence of finite-size effects is to transmute the naive energy level spacings of $1/L$ to spacings $1/NL$. Achieving this with nontrivial flat connections along the orbifolded circle is discussed in e.g. \cite{Horowitz:2001uh}.

The topological volume independence that we discuss seems to be controlling the relation between $\mathcal{N}=4$ super Yang-Mills on $S^3/\mathbb{Z}_p$ and $(2+1)$-dimensional super Yang-Mills on $S^2$, as discussed on the gravity side in \cite{Lin:2005nh} and the field theory side in \cite{Ishiki:2006yr}. An important distinction we draw here from previous work is that the precise pattern of center symmetry breaking/preservation in the gravitational picture is not realized at weak coupling.

It would be fascinating to carry out weakly coupled tests of our proposal for topological volume independence of gauge theory on quotients of simply connected manifolds. A simple case to analyze is that of $(3+1)$-dimensional gauge theory on a lens space. In particular, our arguments (and weakly coupled intuition from an ordinary circle compactification of flat space) suggest that periodic adjoint fermions along the Hopf fiber of the lens space should lead to topological volume independence at weak coupling.  

\section{Extensivity of the Bekenstein-Hawking-Wald entropy}\label{ext}
The Bekenstein-Hawking area law is a universal formula in Einstein gravity that applies to black hole horizons, cosmological horizons, and in a certain sense to spacetime itself. Let us restrict the discussion to black hole horizons and focus on the scaling with area, ignoring the factor of $1/4$. This scaling was explained by Witten for asymptotically large black holes in AdS, since this corresponds to the asymptotically high-temperature limit of the field theory where the entropy should become extensive \cite{Witten:1998zw}. As discussed in the previous section, in this limit the scaling of the field theory entropy with the spatial volume maps directly to the scaling with the area of the horizon in the bulk. The Eguchi-Kawai mechanism, when manifested as the volume-independence of entropy density, seems to be exactly the sort of tool necessary to provide a general mechanism for the area law. But there are several puzzling and ultimately insurmountable features in trying to pinpoint an exact scaling with area purely from the Eguchi-Kawai mechanism (except for large toroidally compactified black branes in AdS). We will instead see that the mechanism explains a more general ``area" law: the extensivity of the Bekenstein-Hawking-Wald entropy.\footnote{The language here and in the literature is very confusing. We refer to the Bekenstein-Hawking entropy as extensive even though it is very famously subextensive. By this we mean extensive in horizon area not volume. Also, the Wald entropy is sometimes referred to as providing subextensive corrections to the Bekenstein-Hawking area law, by which it is meant terms that do not scale with the area of the event horizon. When we refer to the extensivity of the Wald or Bekenstein-Hawking-Wald entropy, we mean the fact that it can be written as an integral of a local quantity over the horizon of the black hole. We will discuss this further below.} Before considering higher curvature corrections, however, let us investigate how the Bekenstein-Hawking area law is at least consistent with the Eguchi-Kawai mechanism, even if not predicted by it.


In AdS$_{d+1}$/CFT$_d$, we may ask why toroidally compactified black branes above the Hawking-Page phase transition have no subextensive piece in their classical entropy. Fixing to a spatial torus, as $\beta \rightarrow 0$ we expect to get an entropy scaling of the conformal field theory as $S \sim V_{d-1} T^{d-1}$. Since the bulk Hawking temperature scales as $T \sim r_h$, this gives $S \sim r_h^{d-1} V_{d-1}$ in bulk variables, which is precisely the Bekenstein-Hawking area law. However, as the temperature is lowered we should generically expect subextensive corrections to the thermal entropy, which would spoil the universal area law in the bulk since $T \sim r_h$ is maintained for black branes at any temperature. However, the Eguchi-Kawai mechanism saves the day, and implies that no such corrections can appear until one undergoes a CSST, whose location can be determined as discussed in section \ref{zeropoint}. This uses the Eguchi-Kawai mechanism to generalize Witten's explanation of the Bekenstein-Hawking area law to all toroidally compactified black branes above the Hawking-Page transition. Of course, if a periodic spin structure is chosen for the fermions along all spatial cycles, then no such transition appears in the gravitational regime and we can explain the area law for arbitrary toroidally compactified black branes. This is just a recap of what was shown more carefully in section \ref{zeropoint}.

What about the Bekenstein-Hawking area law for black hole horizons with curvature, like the spherical or hyperbolic black holes in AdS? Again adopting center-symmetry preservation along the orbifolding cycle (up to any CSST) as our working assumption, we deduce that the entropy density in the field theory is volume-independent in the orbifolding direction. The orbifolding direction is a discrete direction, indexed by an integer $p$ in the previous section. Any potential analytic continuation to complex $p$ is on very shaky ground, but the Bekenstein-Hawking area law for the original spherical or hyperbolic black hole may be understood by analytic continuation from the discrete family of quotiented geometries. This is akin to understanding entanglement entropy through the discrete Renyi family, although there the analytic continuation is on much firmer footing.

If these ideas are correct, then they provide a mechanism for the area law for large black holes with horizon topology $\Sigma$ which dominate the canonical ensemble for some dual field theory on background $\Sigma$. What about small black holes? Here the interpretation in terms of plasma balls in the dual large-$N$ gauge theory may be useful \cite{Aharony:2005bm}. It may then be true that the Eguchi-Kawai mechanism applies to this deconfined plasma ball in a way which maps to the area law in the bulk, as we saw for large black holes above.


\subsection*{Stringy corrections and extensivity of the Wald entropy}
We can ask about subleading order in the 't Hooft coupling $\lambda$, which should correspond to bulk stringy corrections. One way these stringy corrections manifest themselves is as higher-curvature corrections to the bulk Einstein gravity. The Polyakov loop analysis remains the same and continues to indicate center symmetry preservation along $d-1$ cycles. Thus a center-symmetry analysis in the field theory predicts that for any planar/spherical/hyperbolic black holes, the entropy density should be volume-independent in any smooth orbifolding direction.

To check this, we can look at zero-point functions like the entropy density. Since we have higher-curvature corrections we need to use the Wald formula for black hole entropy. For toroidally compactified black branes, the area law is maintained although the coefficient can change. For spherical or hyperbolic black holes, we have corrections to the Bekenstein-Hawking area law which do not scale with the area of the horizon. This seems to be in contradiction with the Eguchi-Kawai mechanism. To address this, let us step back for a moment.

There is a spiritually correct but technically incorrect holographic explanation of the Bekenstein-Hawking area law that is often given. It says that the scaling with area is because there is a holographic dual theory in one lower dimension with the same entropy, and its entropy is scaling with volume as it should be. This captures the holographic spirit, but in general it is technically incorrect as can be seen in many ways. If the area maps to a field theory volume, does the 1/$G_N$ map to temperature? This is of course wrong. Even in the cases where the area does map rigorously to volume, like toroidally compactified black branes, why does the field theory not exhibit any subextensive corrections to its entropy? This we explained within our framework of large-$N$ volume independence. Finally, what about higher curvature corrections? In the bulk the entropy picks up what are sometimes confusingly called ``subextensive corrections to the Bekenstein-Hawking area law" from the Wald entropy formula. This ruins the Bekenstein-Hawking area law. Interpreted as bulk stringy corrections and therefore as corrections in the gauge coupling of a dual field theory, why should going to weaker coupling ruin extensivity?

These issues are clarified by recalling that the Wald entropy is an integral over the event horizon and is therefore extensive. Consider a black hole with metric ansatz
\be
ds^2=-f(r)dt^2+\f{dr^2}{g(r)}+r^2 h_{\mu\nu}d\Sigma^{\mu\nu}\,,
\ee
where $h_{\mu\nu}$ is independent of $r$ and $t$. This does not capture the most general case but will suffice for the argument. The Wald entropy for a general diffeomorphism-invariant higher-curvature theory of gravity with Lagrangian density $\mathcal{L}$ is given as an integral along the horizon
\be
S_W = -2\pi\int d\Sigma \sqrt{h}\, r_h^{d-1} \f{\p \mathcal{L}}{\p R_{\mu\nu\alpha\beta}}\,\epsilon^{\mu\nu}\epsilon^{\alpha\beta}\,,
\ee
where $\epsilon^{\mu\nu}$ is the binormal to the horizon. The corrections implied by the Wald entropy are terms that do not scale as $r_h^{d-1}$, which is the scaling of the Bekenstein-Hawking entropy. But notice that the general theory will still scale with the volume of $\Sigma$: $S_W \sim \text{Vol}(\Sigma)$. This is what we mean by extensivity, which as before can be thought of in terms of quotients of $\Sigma$
\be
\mathcal{M}_{d-1}\longrightarrow \mathcal{M}_{d-1}/\Gamma \qquad\implies \qquad S_W \rightarrow S_W/|\Gamma|\,.
\ee
 In this sense the general Wald entropy -- therefore the entropy in an arbitrary diffeomorphism-invariant theory of classical gravity -- is just as extensive as the Bekenstein-Hawking entropy. For black branes this means that the Wald entropy maintains an area law but in general modifies the Bekenstein-Hawking prefactor  $1/4$. 

To bring this extensivity of curved horizons into clearer focus, consider quantum (subleading in $G_N$, i.e. subleading in $N$) corrections to the Bekenstein-Hawking-Wald entropy. At first order, these are logarithmic in the area of the event horizon:
\be
S_{tot} \sim S_W + \log(S_W)+\dots\,.
\ee
The correction neither scales with the area of the horizon nor with Vol$(\Sigma)$. It is truly subextensive.

This discussion should make clear that the gravity that emerges from our center symmetry analysis is not necessarily Einstein gravity. Nevertheless, it would be fascinating if somehow the stringency of this center symmetry structure necessitated a CFT with an Einstein gravity dual. One way this could occur is by requiring  a sparse higher spin spectrum \cite{Heemskerk:2009pn} -- recently shown to give $c \approx a$ for the anomaly coefficients $c$ and $a$ in four-dimensional CFTs \cite{Afkhami-Jeddi:2016ntf} -- just as it required a sparse spectrum of low-lying states to reproduce the extended range of validity of the general-dimensional Cardy formula. In this spirit, it is encouraging that restoration of a center symmetry plays an important role in deforming higher-spin theory (within which the higher spin fields cannot be made sparse) into ABJ theory (within which they can).

\section{Discussion}\label{disc}

\subsection{Reproducing additional features of AdS gravity}
We have shown that several universal features of AdS gravity can be reproduced with the starting assumption of center symmetry preservation along all but one cycle in a large-$N$ theory (and the suitable generalization of this statement to curved backgrounds as discussed before). However, there are still several features that we would like to explain. 

A powerful technical assumption in the context of reproducing universal features of gravity in AdS$_3$/CFT$_2$ is that of Virasoro vacuum block domination of the four-point function on the sphere. This is expected to be a valid assumption in large-$c$ theories with a sparse light spectrum and sparse low-lying operator-product-expansion (OPE) coefficients. This suggests that it might be implied by our framework. More precisely, consider a four-point function $\langle O_1(\infty), O_2(z) O_3(1) O_4(0)\rangle$, which can be decomposed into representations of the Virasoro algebra (i.e. into Virasoro blocks) by inserting a complete set of states. It is believed that taking $c\rightarrow\infty$ with external and internal operator dimensions scaling with $c$ leads to an exponentiation of the Virasoro block \cite{Belavin:1984vu, Zamolodchikov1987}:
\be
\mathcal{F}(c,h_p,h_i,z)\approx \exp\left(-\f{c}{6}f\left(\f{h_p}{c},\f{h_i}{c},z\right)\right)\,, \qquad i=1,2,3,4\,,
\ee
where $h_p$ is the internal operator dimension. Now taking $z\rightarrow 0$ leads to vacuum block dominance, which means that the $h_p=0$ internal channel gives the largest contribution. This is because all blocks have an expansion around $z=0$ which is dominated by the leading OPE singularity from bringing together $O_2$ and $O_4$:
\be
\mathcal{F}(c,h_p,h_i,z) = z^{h_p-h_2-h_4}(1+\mathcal{O}(z))\,.
\ee
In holographic theories, vacuum block dominance -- like the Cardy formula we discussed in \eqref{zeropoint} -- seems to have an extended range of validity, which in this case means for a finite range of $z$ beyond the asymptotic limit $z\rightarrow 0$. This requires a sparseness bound both on the spectrum of states and on the operator product expansion coefficients. Our framework requires large $c$ to begin with and reproduces a sparse light spectrum as discussed in section \ref{zeropoint}. Data about the OPE coefficients is also accessible in this framework since tree-level Witten diagrams have bulk interactions. Concretely, one may hope to analyze more carefully volume independence for the blocks between the sphere and the torus, possibly using the tools of \cite{Fitzpatrick:2014vua, Hijano:2015rla, Hijano:2015zsa, Hijano:2015qja, Alkalaev:2015wia}. An orthogonal clue that vacuum block dominance may be implied by this framework is a calculation of the entanglement entropy in a heavy microstate on a circle \cite{Asplund:2014coa, Caputa:2014eta, Anous:2016kss}, which gives an answer independent of the size of the circle! 


Accessing some quantity or feature which directly exhibits the smooth, geometric nature of the bulk is another natural goal for this framework. The singularities of \cite{Maldacena:2015iua} are one such feature that indicate a sharp geometric structure.

\subsection{Reducing or blowing up models}
The strong coupling description of holographic theories makes manifest that one can achieve full volume-independence (i.e. preserve center symmetry for all cycle sizes) along directions with periodic (antiperiodic) boundary conditions for fermions (bosons), as long as one direction has the opposite boundary conditions and caps off in the interior. One can then perform a large-$N$ reduction of these theories down to matrix quantum mechanical theories, i.e. $(0+1)$-dimensional theories. For a discussion of the validity of the reduction down to zero size, see appendix \ref{circles}. This captures physics in both confined and deconfined phases. When describing thermal physics in the gravitational limit, there will always be one direction that does not reduce, prohibiting the reduction to a matrix model description, i.e. a $(0+0)$-dimensional theory. (See \cite{Cherman:2012gn} for a discussion of subtleties in dimensionally reducing volume-independent theories.)

Blowing up low-dimensional models is another interesting direction to pursue, especially in light of recent developments in low-dimensional models like the Sachdev-Ye-Kitaev (SYK) model, which captures some features of AdS$_2$ gravity. The addition of flavor to the SYK model \cite{Gross:2016kjj} gives it the necessary ingredient to be blown up into a higher-dimensional model by the methods of \cite{Kovtun:2003hr, Kovtun:2004bz}. (See also \cite{Gu:2016oyy, Berkooz:2016cvq} for a different kind of blow-up.)

\subsection{The necessity of the Eguchi-Kawai mechanism for holographic gauge theories}\label{necek}
I have intermittently referred to the Eguchi-Kawai mechanism as a necessary feature of holographic gauge theories. In a certain sense, this is obviously ridiculous. Center symmetry of e.g. $\mathcal{N}=4$ super Yang-Mills breaks explicitly with the addition of a single fundamental matter field, although we still have a controlled gravitational description of the infrared physics. In this case, what I really mean is that there exists a theory which at large $N$ is equivalent to the one with a single fundamental field, but which has center symmetry at the Lagrangian level. More simply, the fundamental matter decouples at leading order in $N$, so the center symmetry is emergent at infinite $N$. As explored heavily in the literature on large-$N$ volume independence and mentioned in the introduction, orbifold/orientifold dualities in many cases imply an emergent center symmetry at infinite $N$, even when center-breaking matter does not naively decouple \cite{Armoni:2007kd, Shifman:2007kt}. It is this generalized emergent sense in which the Eguchi-Kawai mechanism is necessary. In other words, there is a possibility that center symmetry (whether existing explicitly or emergent) is playing an indispensable role in realizing the precise form of volume independence necessary to admit a gravitational description. Absent conclusive evidence to the contrary, I conjecture this to be the case. It would be nice to have a formalism centered around center symmetry that does not use the crutch of gauge theory, which may be an unnecessary redundancy of description.\footnote{It was pointed out to me by Brian Willett that center symmetry can be discussed in the language of one-form global symmetries, without the need for a Lagrangian, as developed in \cite{Gaiotto:2014kfa}.}

Interesting cases to study, which may teach us about large-$N$ equivalences, are that of the D1-D5 system and of attempts at describing unquenched flavor in AdS/CFT. At the orbifold point, the D1-D5 theory can be thought of as a free symmetric orbifold CFT. It is a gauge theory, but the gauge group is $S_N$ which has a trivial center. Nevertheless, this theory seems to have at least some aspects of large-$N$ volume independence. It realizes the phase structure of gravity, and certain correlators can be written as a sum over images \cite{Balasubramanian:2005qu}. Indeed, the physics of long strings/short strings and sharp transitions controls finite-size effects and an effective circle size which is $1/cL$ instead of the naive $1/L$ (see for example \cite{Aharony:1999ti, Birmingham:2002ph}). The case of unquenched flavor requires keeping $N_f/N$ finite as $N\rightarrow \infty$, which means the flavor does not decouple at leading order in $N$. If there is a smooth gravitational description in AdS (or some similarly warped spacetime), then the nature of finite-size effects should be analyzed.

\subsection{Outlook}
There are many directions to pursue with these ideas in the context of AdS/CFT, only some of which were addressed above. Taking a broader view of the subject, it is clear that holographic dualities which have rules like those of AdS/CFT will have similar volume-independent structure in correlation functions and phase structures. It is remarkable that the mechanism first introduced by Eguchi and Kawai is relevant only in the context of large-$N$ gauge theories, and even then only at leading order in $N$. It is as if it was tailor-made to explain classical gravity, whether within AdS or with some other asymptotia. Indeed, one universal feature of classical gravity we can hang our hats on, robust against changes in asymptotia, is the extensivity of the Bekenstein-Gibbons-Hawking-Wald entropy. The universality of this simple formula only exists at leading order in $G_N$, and we saw that in the context of AdS/CFT it maps to universal volume-independence at leading order in $N$ for certain black holes. It is natural to conjecture that the same mechanism is controlling the entropy for all black holes, although as discussed in the main text this statement should be interpreted with care. The capability of these ideas in addressing classical gravity more generally can be whimsically summarized by the slogan GR = EKR. The extent to which this is a useful and technically accurate perspective beyond AdS/CFT remains to be seen.

\section*{Acknowledgments}
I am greatly indebted to Aleksey Cherman for his many patient explanations of modern developments regarding the Eguchi-Kawai mechanism. I would like to thank Tarek Anous, Aleksey Cherman, and Raghu Mahajan for useful conversations and comments on a draft. I would also like to thank Dionysios Anninos, David Berenstein, William Donnelly, David Gross, Gary Horowitz, Nabil Iqbal, Zohar Komargodski, Don Marolf, Mark Srednicki, Tomonori Ugajin, Mithat Unsal and Brian Willett for useful conversations.

\appendix 
\section{$SL(d,\mathbb{Z})$}\label{app}
In this section we will review some basic points about $SL(d,\mathbb{Z})$, the mapping class group of $\mathbb{T}^d$. When $d$ is even, we will want to consider $PSL(d,\mathbb{Z})$ instead, obtained by quotienting by the center $\{1,-1\}$. For simplicity we will just refer to the group as $SL(d,\mathbb{Z})$ with this distinction implicit. 

\subsection*{Moduli}
Naively, the torus is parameterized by $d$ arbitrary real vectors $V_1,\dots, V_d$ in $d$-dimensional space. However, we can use global rotational invariance to eliminate $\sum_{i=1}^{d-1} i = d(d-1)/2$ of the angles between the vectors, and scale invariance to eliminate a single overall size modulus. The torus now has $d^2-d(d-1)/2-1=(d-1)(d+2)/2$ real moduli. Calling the coordinates $x_1,\dots, x_d$, we have a twist modulus $\theta_{ij}$ between $x_i$ and all $x_j$ with $i<j$, and a size modulus $\theta_{ii}$ for $d-1$ of the cycles $x_i$. Keeping the overall size modulus explicit, we can arrange the moduli in terms of the following lattice vectors:
\begin{equation}
V=\begin{bmatrix}
\vec{V}_1 \\
\vec{V}_2 \\
\vec{V}_3\\
\vdots  \\
\vec{V}_{d-1} \\
\vec{V}_d
\end{bmatrix}=\begin{bmatrix}
\theta_{11}  & \theta_{12}   & \cdots   &\theta_{1,(d-1)}& \theta_{1d}  \\
0  & \theta_{22}    & \cdots &\theta_{2,(d-1)}& \theta_{2d} \\
\vdots  & \vdots & \ddots   & \vdots& \vdots\\
0  &  0 &\cdots     &\theta_{(d-1), (d-1)} & \theta_{(d-1),d}  \\
0  &  0 & \cdots     &0&\theta_{dd}
\end{bmatrix}\,.
\end{equation}

\subsection*{Generators}
In this section we will list four sets of generators of $SL(d,\mathbb{Z})$ and show them to be equivalent. Our first two sets of generators of $SL(d,\mathbb{Z})$ can be written as
\begin{equation}
u_1=\begin{bmatrix}
0  & 1  & 0 & \cdots   & 0  \\
0  & 0  & 1  & \cdots & 0 \\
\vdots  & \vdots & \vdots   & \ddots& \vdots\\
0  &  0 &  0&\cdots     &  1  \\
(-1)^{d+1}  &0  &0 & \cdots     &  0 \\
\end{bmatrix},
\qquad u_2=\begin{bmatrix}
1  & 0  & 0 & \cdots   & 0  \\
1  & 1  & 0  & \cdots & 0 \\
0  &  0 &  1&\cdots     &  0  \\
\vdots  & \vdots & \vdots   & \ddots& \vdots\\
0  &0  &0 & \cdots     &  1 \\
\end{bmatrix}
\end{equation}
or
\begin{equation}
U_1=\begin{bmatrix}
0  & 0 & \cdots &  0& (-1)^{d+1} \\
1  & 0 & \cdots& 0& 0 \\
0  & 1 &\cdots   &  0&  0  \\
\vdots  & \vdots & \ddots   & \vdots&\vdots\\
0  & 0 &\cdots  &   1&  0  \\
\end{bmatrix},
\qquad U_2=\begin{bmatrix}
1  & 1  & 0 & \cdots   & 0  \\
0  & 1  & 0  & \cdots & 0 \\
0  &  0 &  1&\cdots     &  0  \\
\vdots  & \vdots & \vdots   & \ddots& \vdots\\
0  &0  &0 & \cdots     &  1 \\
\end{bmatrix}
\end{equation}
These are $d\times d$ matrices. The small $u$'s can be shown to generate the big $U$'s and vice versa. The relations for e.g. $d=4$ are
\begin{align}
U_1 = u_1^{-1}, \quad U_2=u_1^{-1}u_2u_1^{-2}u_2u_1u_2u_1^{-1}u_2^{-1}u_1 u_2^{-1}u_1 u_2^{-1}u_1^{-1}u_2u_1^{-1}u_2u_1u_2^{-1}u_1^{-1}u_2^{-1}u_1^3\,.
\end{align}
Generating the small $u$'s by the big $U$'s is obtained by swapping $u\leftrightarrow U$.  We will henceforth stick with the big $U$'s. $U_1$ cyclically permutes all the entries of a vector while $U_2$ twists the first vector by an integer amount in the direction of the second vector. The power $d+1$ on the top right element of $U_1$ is necessary to keep det$(U_1)=+1$ and thus stay within $SL(d,\mathbb{Z})$. 

Another set of generators can be given by a simple generalization of the usual $S$ and $T$ generators familiar from $SL(2,\mathbb{Z})$. In this case, we simply have $S_{ij}$ and $T_{ij}$ along any pair of directions $i < j$. Beware the notation: $S_{ij}$ is a $d\times d$ matrix for any given $i,j$, \emph{not} the $\{i,j\}^{\textrm{th}}$ element of a matrix $S$.  Confusingly, $S$ Transposes and $T$ Shears! Better to think of it as $S$ Swaps and $T$ Twists. So we have the elementary row switching (with a minus sign, conventionally placed in the upper triangular part) and upper-triangular row addition (with integer entry) transformations. To see their action more explicitly as matrix multiplication, imagine arranging the lattice vectors row by row into a $d$-dimensional matrix. Then, for example, $T_{25}$ twists direction two by an integer in direction five $\vec{V}_2\rightarrow \vec{V}_2+\vec{V}_5$, and $S_{13}$ transposes lattice vectors as $\vec{V}_1\rightarrow -\vec{V}_3$ and $\vec{V}_3\rightarrow \vec{V}_1$. 

Finally, we can consider as generators the set $T_{ij}$ for generic $i \neq j$. These make up twists in any direction. These include the upper-triangular $T_{ij}$ from the previous section and generate the swaps $S_{ij}=T_{ji}(T_{ij})^{-1}T_{ji}$. Conversely, these lower-triangular twists can be generated by the previous set of generators as $T_{ji} = (T_{ij})(S_{ij})(T_{ij})$. The set of swaps and upper-twists can also generate $U_1$ and  $U_2$ as $U_1= (S_{12})(S_{23})\cdots (S_{d-1,d})$ and $U_2=T_{12}$.

\section{Four-point function sample calculation}\label{fourpoint}
\begin{figure}
\centering
\includegraphics[scale=0.3]{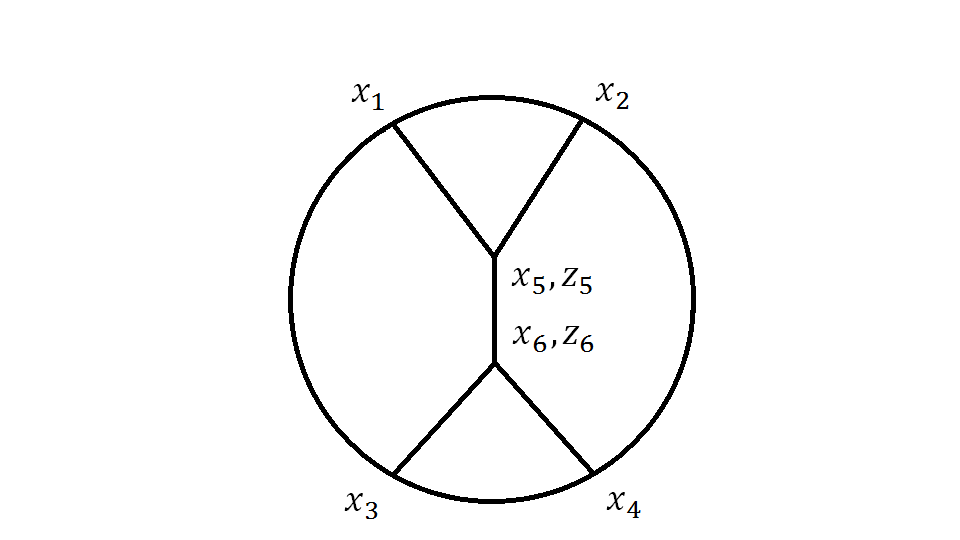}
\caption{\label{wittendiagfour} Left: A tree-level Witten diagram, which contributes at leading order in $N$ to the four-point function. It is constructed out of $M=4$ bulk-to-boundary propagators and $n=1$ bulk-to-bulk propagator. Since it is a contribution at tree level, there are $2=n+1$ interaction vertices. There are more diagrams contributing at this order, including the one with the four bulk-to-boundary propagators meeting at a single interaction vertex in the interior.}
\end{figure}

Here we calculate the tree-level contribution to the four-point function illustrated in figure \ref{wittendiagfour}. We will calculate it in an AdS background where one direction has size $L$ and another AdS background where the same direction has size $JL$ for $J\in \mathbb{Z}^+$. We will suppress all other directions.

We first calculate the correlator in size $JL$. We have
\be
\langle O(x_1)\cdots O(x_4)\rangle_{JL} = \int dz_5 \,dz_6\int_0^{JL} dx_5 \int_0^{JL} dx_6\, K(x_1-x_5)\cdots K(x_4-x_6)G(x_5-x_6)\,.
\ee
Fourier transforming gives
\be
\int \hspace{-1mm}dz_5 \,dz_6\int_0^{JL} \hspace{-2mm}dx_5 \int_0^{JL} \hspace{-2mm}dx_6  \sum_{s_i\in \mathbb{Z}} e^{\f{2\pi i}{JL}(s_1(x_1-x_5)+\dots+s_4(x_4-x_6)+s_5(x_5-x_6))}K(s_1/JL)\cdots K(s_4/JL)G(s_5/JL),
\ee
where $i=1,\dots,5$. Evaluating the $x_5$ and $x_6$ integrals gives
\be
\int dz_5 \,dz_6 \,J^2L^2\sum_{s_i\in \mathbb{Z}}  \delta_{s_1+s_2,s_5}\delta_{s_3+s_4,-s_5}e^{\f{2\pi i }{JL}(s_1x_1+\dots + s_4x_4)}K(s_1/JL)\cdots K(s_4/JL)G(s_5/JL)\,.
\ee
Performing the sum over $s_5$ gives
\be
\int dz_5\, dz_6\sum_{s_i\in \mathbb{Z}}J^2L^2\delta_{s_1+s_2+s_3+s_4,0} e^{\f{2\pi i }{JL}(s_1x_1+\dots + s_4x_4)}K(s_1/JL)\cdots K(s_4/JL)G((s_1+s_2)/JL)\,,
\ee
where now $i=1,\dots,4$. To obtain the momentum space correlator, we do the discrete transform with respect to the variables $x_i$. Recall that the discrete transforms in finite size look like
\be
f(x)=\sum_{n\in \mathbb{Z}} e^{2\pi i n x/L}f(n/L)\implies f(n/L)=\f{1}{L} \int_0^L dx\,e^{-2\pi i n x/L}f(x)\,.
\ee
So we have
\begin{align}
\langle O(n_1/L)\dots O(n_4/L)\rangle_{JL} = \f{J^2L^2}{J^4L^4}\int_0^{JL} dx_1 \dots dx_4 &\int dz_5 \,dz_6 \\
\sum_{s_i\in \mathbb{Z}}e^{\f{2\pi i}{JL}(s_1x_1+\dots s_4x_4)+\f{2\pi i}{L}(n_1x_1+\dots+n_4x_4)}&K(s_1/JL)\cdots K(s_4/JL)G((s_1+s_2)/JL)\,.
\end{align}
Evaluating the integrals and then the sum gives
\be\label{final}
J^2L^2 \int dz_5 \,dz_6 K(n_1/L)\dots K(n_4/L)G((n_1+n_2)/L)\delta_{n_1+n_2+n_3+n_4,0}\,.
\ee

Now we consider the correlator in size $L$, where we replace the bulk-to-bulk propagator and the bulk-to-boundary propagators with those of size $JL$ by the method of images:
\begin{align}
\langle O(x_1)\cdots O(x_4)\rangle_L = \int dz_5 \,dz_6\int_0^L dx_5 \int_0^L dx_6  \sum_{n_i=0}^{J-1} K(x_1+n_1L-x_5)K(x_2+n_2L-x_5)\\
\times K(x_3+n_3L-x_6)K(x_4+n_4L-x_6)G(x_5+n_5L-x_6)\,,
\end{align}
where $i=1,\dots,5$. We Fourier transform these propagators, which have periodicity $JL$, to get
\begin{align}
\int dz_5 &\,dz_6\int_0^L dx_5 \int_0^L dx_6 \sum_{n_i=0}^{J-1} \sum_{n_i'=-\infty}^{\infty} \\
&e^{\f{2\pi i }{JL}\left(n_1'(x_1+n_1L-x_5)+\dots+n_4'(x_4+n_4L-x_6)+n_5'(x_5+n_5L-x_6)\right)}K(n_1'/JL)\cdots K(n_4'/JL)G(n_5'/JL)\,.
\end{align}
Switching the two sums and evaluating the sums over $n_i$ gives
\begin{align}
\int dz_5 &\,dz_6\int_0^L dx_5 \int_0^L dx_6 \sum_{n_i'=-\infty}^{\infty} \delta_{n_1',Js_1}\cdots\delta_{n_4',Js_4}\delta_{n_5',Js_5} J^5\\
&e^{\f{2\pi i }{JL}\left(n_1'(x_1-x_5)+\dots+n_4'(x_4-x_6)+n_5'(x_5-x_6)\right)}K(n_1'/JL)\cdots K(n_4'/JL)G(n_5'/JL)
\end{align}
for arbitrary integer $s_i$. Evaluating the sums over $n_i'$ gives
\begin{align}
\int dz_5 &\,dz_6\int_0^L dx_5 \int_0^L dx_6 \,J^5\sum_{s_i=-\infty}^{\infty}\\
&e^{\f{2\pi i }{L}\left(s_1(x_1-x_5)+\dots+s_4(x_4-x_6)+s_5(x_5-x_6)\right)}K(s_1/L)\cdots K(s_4/L)G(s_5/L)\,.
\end{align}
Performing the $x_5$ and $x_6$ integrals gives
\be
\int  dz_5 \,dz_6\, J^5L^2\sum_{s_i=-\infty}^{\infty}\delta_{s_1+s_2,s_5}\delta_{s_3+s_4,-s_5}e^{\f{2\pi i }{L}\left(s_1 x_1+\dots +s_4x_4+s_5x_5\right)}K(s_1/L)\cdots K(s_4/L)G(s_5/L)\,.
\ee
Performing the sum over $s_5$ gives
\be
\int  dz_5 dz_6\,J^5 L^2 \sum_{s_i=-\infty}^{\infty}\delta_{s_1+s_2+s_3+s_4,0}\,e^{\f{2\pi i }{L}\left(s_1 x_1+\dots +s_4x_4\right)}K(s_1/L)\cdots K(s_4/L)G((s_1+s_2)/L)\,.
\ee
where now $i=1,\dots,4$. The momentum-space correlator is obtained as before:
\begin{align}
\hspace{-2mm}\langle O(n_1/L)\cdots O(n_4/L)\rangle_L &= \f{1}{L^4}\int_0^Ldx_1\dots dx_4 \,\langle O(x_1)\cdots O(x_4)\rangle_L\, \,e^{-\f{2\pi i}{L} (n_1 x_1+\dots+n_4x_4)}\\
&=J^5L^2\int dz_5\, dz_6K(n_1/L)\cdots K(n_4/L) G((n_1+n_2)/L)\delta_{n_1+n_2+n_3+n_4,0} \,.
\end{align}
This is our final answer for the correlator in size $L$. Comparing this answer to \eqref{final} gives us
\be
\langle O(n_1/L)\cdots O(n_4/L)\rangle_L = J^3\langle O(n_1/L)\cdots O(n_4/L)\rangle_{JL}
\ee
 as predicted by \eqref{toprove}.

This calculation should make clear that \eqref{toprove} is correct diagram-by-diagram in the bulk. Moreover, any bulk-to-bulk propagator with momenta that need to be integrated over, as would be the case for loop diagrams, would ruin this structure. This is expected since the presence of such propagators signals a subleading-in-$N$ Witten diagram, for which volume-independence does not apply.

\section{Validity of gravitational description}\label{circles}
For our gravitational description to be valid, we need to deal with smooth geometries and keep cycle sizes larger than string scale. The first criterion is simply because singularities are not well-described within gravity. The second criterion is because stringy excitations (e.g. strings that wrap the cycles) will become important for cycles that are string scale. In this case, one needs to T-dualize along the small cycle to blow it up. The language here is a bit confusing, as T-dualizing takes us from a valid IIB gravity description to a valid IIA gravity description, but we are concerned with maintaining a valid gravity description in the same frame throughout.

Maintaining validity of the gravitational description depends on the periodicity conditions chosen for the matter fields. To be very concrete, let us consider the duality between Type IIB string theory in AdS$_5\times S^5$ and (3+1)-dimensional $\mathcal{N}=4$ super Yang-Mills compactified on a spatial three-torus of cycle lengths $L_i$. First consider the case where the matter fields are given supersymmetry-preserving boundary conditions along the spatial cycles. In this case the ground state geometry is given by the Poincar\'e patch with periodic identifications in the spatial directions. But this means that the cycles become arbitrarily small as the horizon is approached, necessitating a breakdown of the IIB gravity description. This was the case analyzed in \cite{Poppitz:2010bt}. However, finite temperature is different and necessitates a discussion of the order of limits taken. The Euclidean geometry is that of the black brane:
\be
ds^2=\f{f(r)dt_E^2}{\ell_{\text{AdS}}^2}+\ell_{\text{AdS}}^2\f{dr^2}{r^2}+\f{r^2}{\ell_{\text{AdS}}^2} d\phi_id\phi^i\,,\qquad f(r) = r^2(1-(r_h/r))^4, \qquad \beta = \f{\pi\ell_{\text{AdS}}^2}{r_h}\,,
\ee
where $r\geq r_h$, the $S^5$ is suppressed, and $t_E \sim t_E+\beta$ gives the inverse temperature. The minimum proper size of a given cycle $\phi_i$ occurs at $r_h$. This size must be bigger than the string scale $\ell_s$, which gives us the condition
\be
\f{r_h L_i}{\ell_{\text{AdS}}} \gg \ell_s \implies \f{\ell_{\text{AdS}}}{\ell_s}\sim\lambda^{1/4}\gg \f{\beta}{L_i}\,.
\ee
Here we have brought in the 't Hooft coupling $\lambda$. We see that we can make $L_i$ arbitrarily small and maintain validity of the gravitational description as long as we take $\lambda \rightarrow \infty$ first. In other words, we do not scale any cycle sizes with the 't Hooft coupling as we take the strong coupling limit $\lambda\rightarrow \infty$.

The case we were more preoccupied with in the text, especially in section \ref{zeropoint}, is that of modular $U_1$-invariant boundary conditions. This means supersymmetry-breaking boundary conditions along all cycles. As we saw, this implies that when a cycle size is the smallest, it caps off in the interior. The geometry that dominates is either the black brane or the AdS soliton, whose Euclidean continuations are identical. The condition above therefore generalizes to 
\be
\f{\ell_{\text{AdS}}}{\ell_s} \gg \f{L_{\mu,\text{min}}}{L_i}\,,
\ee
where $L_{\mu,\text{min}}$ is the minimum cycle size. By definition we have $\f{L_{\mu,\text{min}}}{L_i}<1$, so this condition is satisfied trivially. Any time a cycle tries to become substringy, it instead caps off. 

Mixed boundary conditions which preserve some subgroup of the full modular $U_1$ invariance are analyzed similarly. The final conclusion is that the gravitational description will remain valid for all cycle sizes as long as at least one cycle has supersymmetry-breaking boundary conditions and remains finite sized in the CFT. The one caveat is that any supersymmetry-preserving cycles are not taken to zero size as an inverse power of the 't Hooft coupling $\lambda$.

\footnotesize
\bibliographystyle{apsrev4-1long}
\bibliography{LargeNvolumeIndependenceBIB}

\end{document}